\documentclass[aps,prd,twocolumn,showpacs,superscriptaddress,groupedaddress]{revtex4} 
\usepackage{amsmath,amssymb,amsfonts,amsthm,bm,mathalfa,array}    
\usepackage{epsfig,psfrag,graphicx}
\usepackage{float}
\usepackage{stackengine}
\newcommand\xrowht[2][0]{\addstackgap[.5\dimexpr#2\relax]{\vphantom{#1}}}
 
\begin{document}
\title{Shifted $\bm{\mu}$-hybrid inflation, gravitino dark matter, and\\ observable gravity waves}
\date{\today}
\author{George Lazarides}
\email[E-mail: ]{lazaride@eng.auth.gr}
\affiliation{School of Electrical and Computer Engineering, Faculty of Engineering, Aristotle University of Thessaloniki, Thessaloniki 54124, Greece}
\author{Mansoor Ur Rehman}
\email[E-mail: ]{mansoor@qau.edu.pk}
\affiliation{Department of Physics, Quaid-i-Azam University, Islamabad 45320, Pakistan}
\author{Qaisar Shafi}
\email[E-mail: ]{shafi@bartol.udel.edu}
\affiliation{Bartol Research Institute, Department of Physics and Astronomy, University of Delaware, Newark, DE 19716, USA}
\author{Fariha K. Vardag}
\email[E-mail: ]{f.vardag@gmail.com}
\affiliation{Department of Physics, Quaid-i-Azam University, Islamabad 45320, Pakistan}
\begin{abstract}

We investigate supersymmetric hybrid inflation in a realistic model based on the gauge symmetry 
$SU(4)_c \times SU(2)_L \times SU(2)_R$. The minimal supersymmetric standard model (MSSM) $\mu$ term arises, following Dvali, Lazarides, and Shafi, from the coupling of the MSSM electroweak doublets to a gauge singlet superfield which plays an essential role in inflation. The primordial monopoles are inflated away by arranging that the $SU(4)_c \times SU(2)_L \times SU(2)_R$ symmetry is broken along the inflationary trajectory.
The interplay between the (above) $\mu$ coupling, the gravitino mass, and the reheating following inflation is discussed in detail. We explore regions of the parameter space that yield gravitino dark matter and observable gravity waves with the tensor-to-scalar ratio $r \sim 10^{-4}-10^{-3}$.

\end{abstract}
\maketitle
\section{\label{Intro}Introduction}
In its simplest form supersymmetric (SUSY) hybrid inflation \cite{Dvali:1994ms,Copeland:1994vg} is associated with a gauge symmetry breaking $G \rightarrow H$, and it employs a minimal renormalizable superpotential $W$ and a canonical K\"ahler potential $K$. Radiative corrections and soft SUSY breaking terms together play an essential role \cite{Senoguz:2004vu,Rehman:2009nq,Pallis:2013dxa,Buchmuller:2014epa} in the inflationary potential that yields a scalar spectral index in full agreement with the Planck data \cite{Akrami:2018odb}. In this minimal model the symmetry breaking $G\rightarrow H$ occurs at the end of inflation, and the symmetry breaking scale is predicted to be of the order of $(2-3) \times 10^{15}{\rm\ GeV}$ \cite{Dvali:1994ms,Senoguz:2004vu,Rehman:2009nq,Pallis:2013dxa,Buchmuller:2014epa}. One simple extension of this minimal model retains a minimal $W$ but invokes a nonminimal $K$ \cite{BasteroGil:2006cm}, such that the correct scalar spectral index is obtained without invoking the soft SUSY breaking terms. Nonminimal K\"ahler potentials are also used to realize symmetry breaking scales comparable to the grand unified symmetry 
(GUT) scale $M_{\text{GUT}}$ ($\sim2 \times 10^{16}{\rm\ GeV}$) \cite{urRehman:2006hu}, and to predict possibly observable gravity waves \cite{Rehman:2010wm,Civiletti:2014bca}.

If the symmetry breaking $G\rightarrow H$ produces topological defects such as magnetic monopoles, a more careful approach is required in order to circumvent the primordial monopole problem. The first such example is provided by the so-called ‘shifted-hybrid inflation’ 
\cite{Jeannerot:2000sv,Jeannerot:2001xd}, in which the monopole producing Higgs field actively participates in inflation such that, during inflation, $G$ is broken to $H$ and the monopoles are inflated away.

In this paper we explore inflation and reheating in the framework of the gauge symmetry $SU(4)_c \times SU(2)_L \times SU(2)_R$
($G_{\text{4-2-2}}$) \cite{Pati:1974yy}. A SUSY model based on this symmetry including hybrid inflation was first explored in Ref.~\cite{King:1997ia}. However, the primordial monopole problem was not resolved, but it was subsequently addressed and successfully rectified in Ref.~\cite{Jeannerot:2000sv} based on shifted hybrid inflation. In the model proposed here, we employ the mechanism invented in Refs.~\cite{King:1997ia,Dvali:1997uq} for generating the MSSM $\mu$ term, and we exploit shifted hybrid inflation to overcome the monopole problem. We implement this scenario using both minimal and nonminimal K\"ahler potentials, and address in both cases important issues related to the gravitino problem  \cite{Ellis:1984eq}. For a discussion of leptogenesis via right-handed neutrinos in models where the dominant inflaton decay channel yields higgsinos, see Ref.~\cite{Lazarides:1998qx}. 

The plan of the paper is as follows: In Sec.~\ref{Mod}, we present the SUSY $G_{\text{4-2-2}}$ model including the superfields, their charge assignments, and the superpotential which respects a $U(1)_R$ symmetry. In Sec.~\ref{inf}, the inflationary setup is described. This includes the scalar potential for global SUSY as well as the one including supergravity (SUGRA). The shifted $\mu$-hybrid inflation ($\mu$HI) scenario with minimal K\"ahler potential and its compatibility with the gravitino constraint \cite{Okada:2015vka} is studied in Sec.~\ref{min}. The analysis is extended by employing a nonminimal K\"ahler potential in Sec.~\ref{NM}, discussing again the gravitino problem and the bounds it imposes on reheat temperature, and focusing on solutions with observable gravity waves. Our conclusions are summarized in Sec.~\ref{conclusion}.
\section{\label{Mod}The supersymmetric $\bm{SU(4)_c\times SU(2)_L \times SU(2)_R}$ model}
The matter and Higgs superfields of the SUSY $G_{\text{4-2-2}}$ model with their representations, transformations under $G_{\text{4-2-2}}$, decompositions under $G_{SM}$, and charge assignments are shown in Table~\ref{assign1}. The matter superfields $F_i$ and $F_i^c$ belong in the following representations of $G_{\text{4-2-2}}$:
\begin{align}
F_i=(4, 2,1)\equiv
  \left( {\begin{array}{cccc}
   u_{ir} &  u_{ig}  &  u_{ib}  & \nu_{il} \\
   d_{ir} &  d_{ig}  &  d_{ib}  & e_{il} \\
  \end{array} } \right),\nonumber \\
   F^c_i\!=(\overline 4, 1, 2)\equiv 
  \left( {\begin{array}{cccc}
   u^c_{ir} &  u^c_{ig}  &  u^c_{ib}  & \nu^c_{il} \\
   d^c_{ir} &  d^c_{ig}  &  d^c_{ib}  & e^c_{il} \\
    \end{array} } \right),
\end{align}
  where the index \textit{i}(=\ 1, 2, 3) denotes the three families of quarks and leptons, and the subscripts $r,\ g,\ b,\ l$ are the four colors in the model, namely red, green, blue, and lilac. 
  The GUT Higgs superfields $H^c$ and $\overline{H^c}$ are represented as follows:
 \begin{align}
H^c\,=(\overline4, 1, 2)\equiv 
  \left( {\begin{array}{cccc}
   u^c_{Hr} &  u^c_{Hg}  &  u^c_{Hb}  & \nu^c_{Hl}\\
   d^c_{Hr} &  d^c_{Hg}  &  d^c_{Hb}  & e^c_{Hl} \\
  \end{array} } \right),\nonumber \\
  \overline {H^c}\!=(4, 1, 2)\equiv 
  \left( {\begin{array}{cccc}
   \overline {u^c_{Hr}} &  \overline{u^c_{Hg}}  & \overline{u^c_{Hb}}  & \overline{ \nu^c_{Hl}}\\\
   \overline {d^c_{Hr}} & \overline {d^c_{Hg}}  & \overline{d^c_{Hb}}  &\overline {e^c_{Hl}} \\
    \end{array} } \right),
\end{align}
and acquire nonzero vacuum expectation values (vevs) along the right-handed sneutrino directions, that is $|\langle \nu^c_{Hl}\rangle|=|\langle \overline{ \nu^c_{Hl}}\  \rangle |=v\neq0$, to break the 
$G_{\text{4-2-2}}$ gauge symmetry to the standard model (SM) gauge symmetry \big($G_{SM}=SU(3)_c\times SU(2)_L\times U(1)_Y$\big), around the GUT scale ($\sim 2\times10^{16}{\rm\ GeV}$), while preserving low scale SUSY
\cite{Shafi:1998yy}. The electroweak  breaking is triggered by the electroweak Higgs doublets, $h_u$ and $h_d$, which reside in the bidoublet Higgs superfield $h$ represented as follows:
\begin{equation}
h=(1, 2, 2)\equiv (h_u \ \ h_d)= \left( {\begin{array}{cc}
    h^+_u &   h^0_d   \\
  h^0_u &     h^-_d   \\
  \end{array} } \right).\\
\end{equation}
Note that such doublets can remain light because of appropriate discrete symmetries 
\cite{lightdoublets}. 
A gauge singlet chiral superfield $S=(1, 1, 1)$ is introduced, which triggers the breaking of 
$G_{\text{4-2-2}}$ and whose scalar component plays the role of the inflaton. A sextet Higgs superfield $G = (6, 1, 1)$, which under the SM splits into the color-triplet Higgs superfields $g = (3, 1, -1/3)$ and $g^c = (\overline 3, 1, 1/3)$, is introduced to provide superheavy masses to the color-triplet pair $d^c_H$ and $\overline{d^c_H}$ \cite{King:1997ia}. 
 \begin{table}
\caption{\label{assign1} Superfields together with their decomposition under the SM and their $R$ charge.}
\begin{ruledtabular}
\begin{tabular}{ccccc}
\xrowht{10pt}
Superfields&$4_c\times 2_L \times2_R$ & $3_c\times 2_L \times1_Y$     &$q(R)$  \\
\hline  
  \hline
  \xrowht{10pt}
$F_i$& $({4,\ 2,\ 1})$  & $Q_{ia} ({ 3,\ 2},\ \ \ 1/6)$&1 \\
\xrowht{10pt}
&  & $L_i ({ 1,\ 2},\ -1/2)$&  \\
\hline
\xrowht{10pt}
$F^c_i$& $({\overline{4},\ 1,\ 2})$ & $u^c_{ia}({ \overline{3},\ 1},\ -2/3)$&1  \\
\xrowht{10pt}
&   & $d^c_{ia}({ \overline{3},\ 1},\ \ \ 1/3)$&  \\
\xrowht{10pt}
&  &$\nu^c_i \ ({ 1,\ 1},\ \ 0)$&\\
\xrowht{10pt}
&   & $e^c_i \ ({ 1,\ 1},\ \ 1)$&\\
\hline
\hline
\xrowht{10pt}
 $H^c$& $ ({\overline{4},\ 1,\ 2})$ & $u^c_{Ha} ({ \overline{3},\ 1},\ -2/3)$&0\\
 \xrowht{10pt}
&  & $d^c_{Ha}({ \overline{3},\ 1},\ \ \ 1/3)$&\\
\xrowht{10pt}
&  &$\nu^c_H\  ({ 1,\ 1},\ \  0)$&\\
\xrowht{10pt}
&  & $e^c_H \ ({ 1,\ 1},\ 1)$&\\
\hline
\xrowht{10pt}
$\overline {H^c}$& $ ({4,\ 1,\ 2})$ & $\overline {u^c_{Ha}} ({ 3,\ 1},\ \ \ 2/3)$&0  \\
\xrowht{10pt}
&  & $\overline {d^c_{Ha}} ({3,\ 1},\ \ \ -1/3)$& \\
\xrowht{10pt}
& & $\overline {\nu^c_H}\  ({ 1,\ 1},\ \  0)$& \\
\xrowht{10pt}
& &$\overline {e^c_H}\  ({ 1,\ 1},\  -1)$& \\
\hline
\xrowht{10pt}
$ S$& $ ({1,\ 1,\ 1})$ & $ S({ 1,\ 1},\ \ \ 0)$&2 \\
\hline
\xrowht{10pt}
$ G$& $ ({6,\ 1,\ 1})$ & $g_{a} ({3,\ 1},\ -1/3)  $&2 \\
\xrowht{10pt}
$ $&  &$ g^c_{a} ({\overline{3},\ 1},\  \  \ 1/3)$& \\
\hline
\xrowht{10pt}
 $h$&$({1,\ 2,\ 2})$ & $h_u\  ({1,\ 2},\ \ \ 1/2)$&0 \\
 \xrowht{10pt}
&  &$h_d\ ({ 1,\ 2},\ -1/2)$&  \\
\end{tabular}
\end{ruledtabular}
\end{table}

The main part of the superpotential of our model that is compatible with $G_{\text{4-2-2}}$ and the 
R-symmetry $U(1)_R$ is given by \begin{align}\label{SP1}
W & = \kappa S (\overline{H^c}H^c-M^2)+\lambda Sh^2 \nonumber \\
&-S \left( \beta_1 \frac {(\overline{H^c} H^c)^2}{\Lambda^2}
+\beta_2 \frac {(\overline{H^c})^4}{\Lambda^2}+\beta_3 \frac {(H^c)^4}{\Lambda^2} \right)\nonumber \\ 
&+ \lambda_{ij}F^c_iF_jh+\gamma_{ij}\frac{\overline{H^c}\ \overline{H^c}}{\Lambda}F^c_iF^c_j\nonumber  \\ &+a\,GH^cH^c+b\,G\overline{H^c}\ \overline{H^c},
\end{align}
where $\kappa,\ \lambda,\ \beta_{1,2,3},\  \lambda_{ij},\  \gamma_{ij},\  a, \text{ and } b$ are real and positive dimensionless couplings and $M$ is a mass parameter of the order of $M_{\rm GUT}$. We assume the superheavy scale $\Lambda$ to be in the range $10^{16}{\rm\ GeV}$$\lesssim \Lambda \lesssim m_P$, where $m_P$ denotes the reduced Planck scale ($2.4 \times 10^{18}{\rm\ GeV}$). The first three terms in the superpotential are of the standard $\mu$HI case as discussed in Refs.~\cite{Okada:2015vka, Rehman:2017gkm}. The first two and the fourth term characterize the `shifted case' by providing additional inflationary tracks to avoid the monopole problem. The third term $\lambda S h_u h_d$ yields the effective $\mu$ term. Indeed assuming gravity-mediated SUSY breaking \cite{Chamseddine:1982jx,Linde:1997sj}, the scalar component of $S$ acquires a nonzero vev proportional to the gravitino mass $m_{3/2}$ and generates a $\mu$ term with $\mu=-\lambda m_{3/2}/\kappa$, thereby resolving the MSSM $\mu$ problem \cite{Dvali:1997uq}.  The $\lambda_{ij}$-terms contain the Yukawa couplings, and hence provides masses for fermions. The $\gamma_{ij}$-terms yield large right-handed neutrino  masses, needed for the see-saw mechanism. The other possible couplings similar to $\gamma_{ij}$-terms which are allowed by the symmetries are
$FFH^cH^c$, $FF\overline{H^c}\  \overline{H^c}$, and $F^cF^cH^cH^c$. The last two terms in the superpotential involving the sextuplet superfield $G $ are included to provide superheavy masses to $d^c_H$ and $\overline{d^c_H}$. 
 
  This model can be embedded in a realistic supersymmetric $SO(10)$ GUT model along the same 
lines as in Ref.~\cite{Kyae:2005vg}, where the matter superfields $F$ and $F^c$ are 
combined in a $16$, the Higgs superfield $H^c$ together with a (4,2,1) in a 
$16_H$, and the $\overline{H^c}$ together with a ($\overline{4}$,2,1) in 
a $\overline{16_H}$. The bidoublet $h$ together with a sextet will reside 
in a $10_h$. An additional Higgs superfield such as 210 or 54 will be needed to 
break $SO(10)$ to $G_{\text{4-2-2}}$.

 It is important to mention here that the matter-parity symmetry $\mathbb Z_2^{mp}$, which is usually invoked to forbid rapid proton decay operators at renormalizable level, is contained in $U(1)_{R}$ as a subgroup. The superpotential $W$ is invariant under $\mathbb Z_2^{mp}$ and this symmetry remains unbroken. There is no domain wall problem and the lightest SUSY particle (LSP) is stable and consequently a plausible candidate for dark matter (DM). 
\section{\label{inf}$\bm\mu$-hybrid inflation in $\bm{SU(4)_c\times SU(2)_L \times SU(2)_R}$}
The relevant part of the superpotential for shifted $\mu$HI contains the terms
\begin{equation}\label{SP}
\delta W= \kappa S(\overline{H^c}H^c-M^2)+\lambda S h^2-\xi \frac {\kappa S(\overline {H^c}H^c)^2}{M^2},
\end{equation}
where $\xi=\beta_1 M^2/\kappa \Lambda^2$ is a dimensionless parameter. We ignore the $\beta_{2,3}$-terms in our future discussions as they become irrelevant in the $D$-flat direction, that is the direction where the $D$-term contributions vanish (i.e. with $|\nu_H^c| = |\overline{\nu_H^c}|$ and all other components zero). For simplicity, the superfields and their scalar components will be denoted by the same notation. 

The global SUSY minimum obtained from Eq.~(\ref{SP}) is given as
\begin{equation}
\langle S\rangle\!=\!0,\ \ \ \ \langle h \rangle\!=\!0,\ \ \ \ v^2\!=\!\langle \overline{H^c} H^c\rangle\!=\! \frac{M^2}{2\xi}(1\!\pm\sqrt{1\!-\!4\xi}), 
\end{equation}
which requires that $\xi\leq 1/4$ for real values of $v$. Note that, for $\xi>1/4$, the global SUSY vacuum lies at complex values of the fields $H^c$, $\overline{H^c}$, but we will not consider this 
case in this paper. 

\begin{figure*}
\includegraphics[scale=0.4]{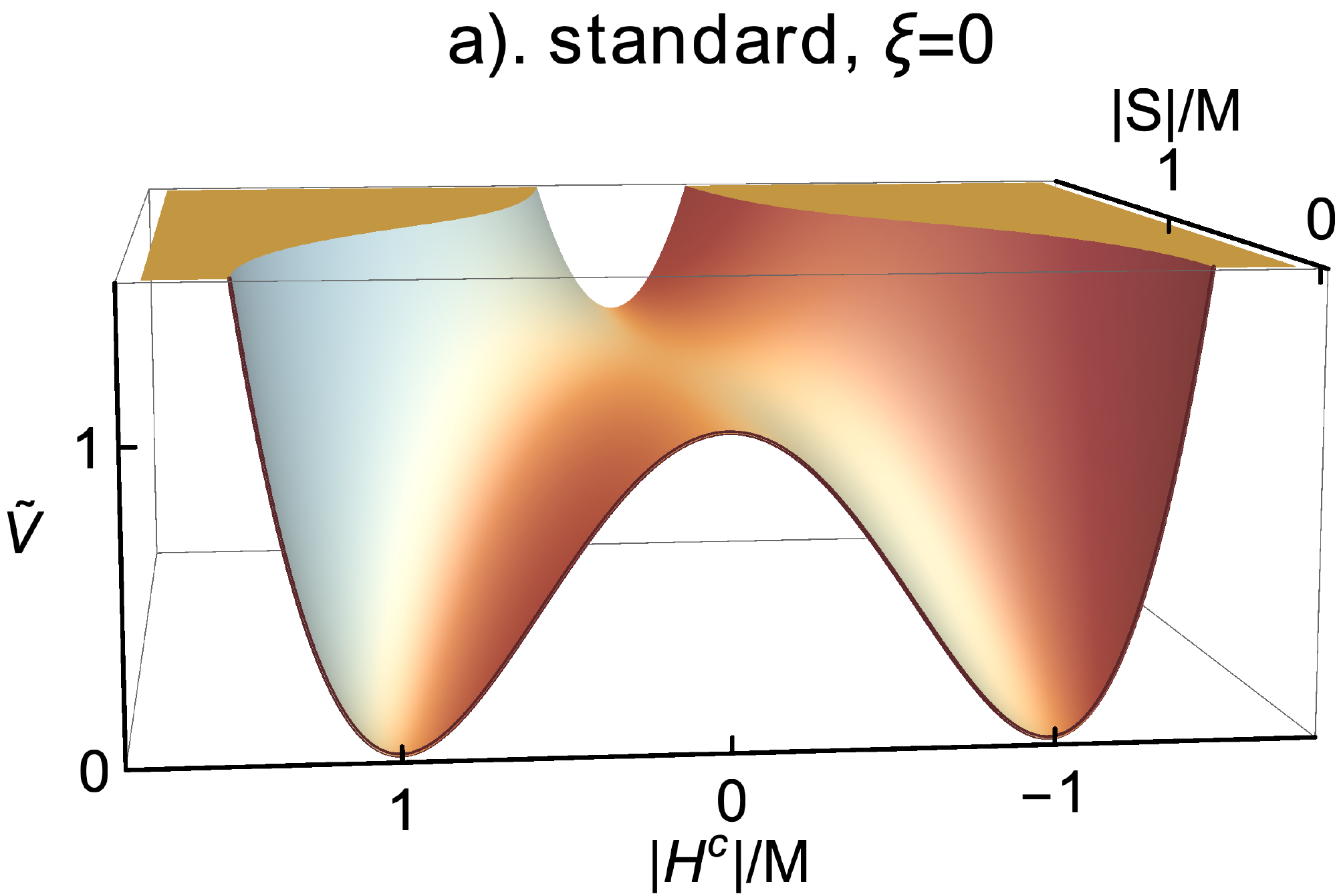}
\includegraphics[scale=0.4]{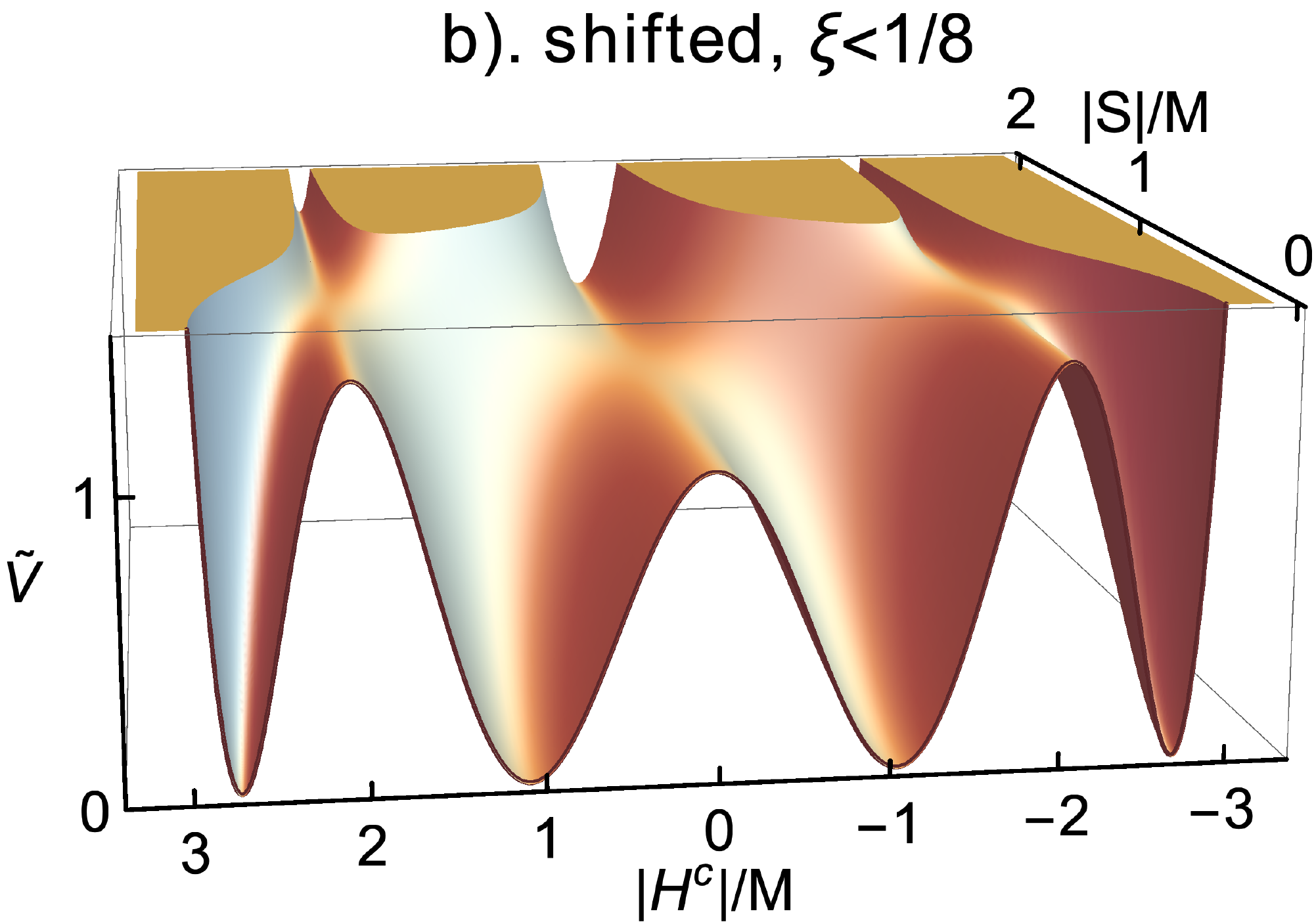}
\par
 \vspace{1.75cm}
\includegraphics[scale=0.4]{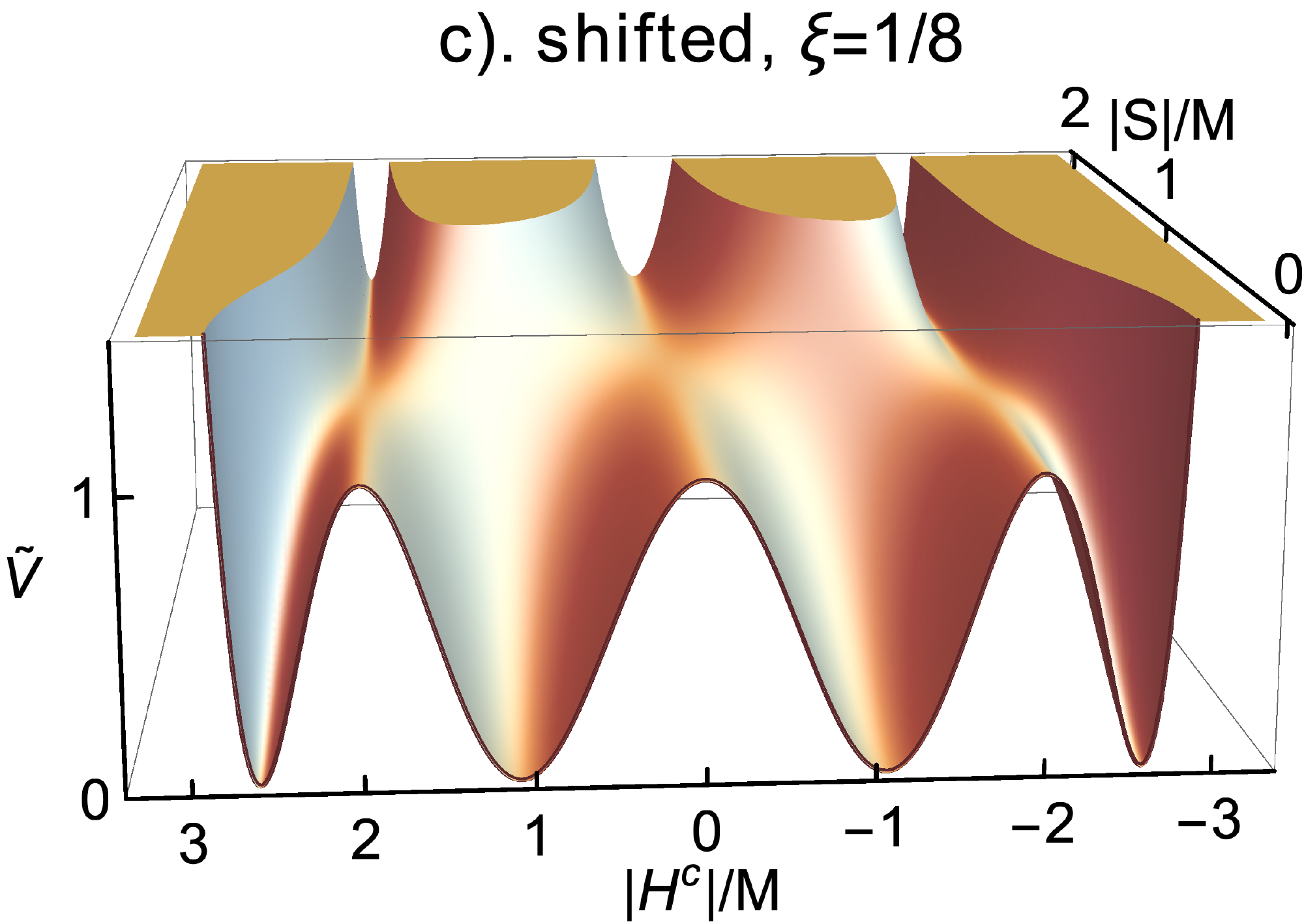}
\includegraphics[scale=0.4]{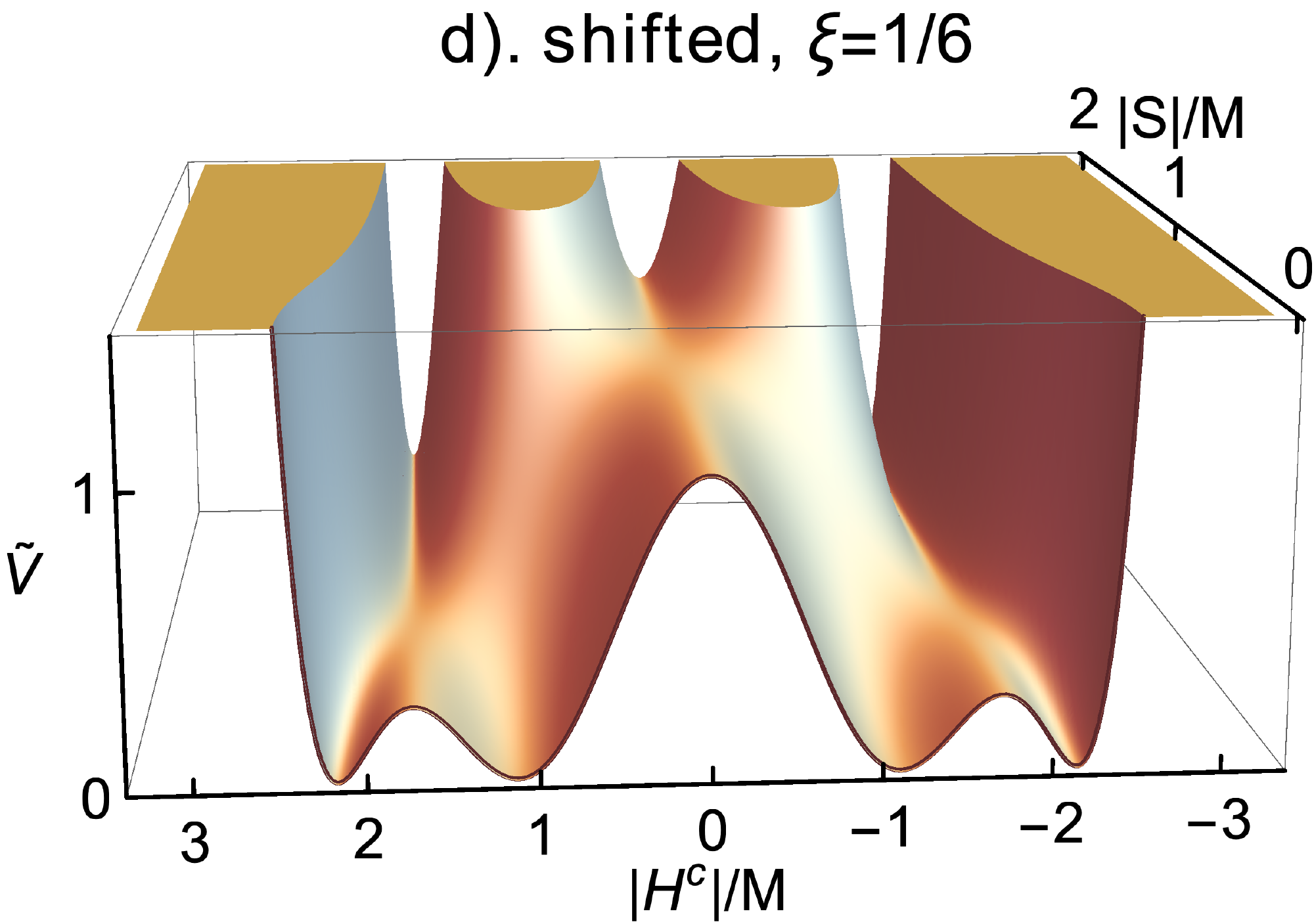}
\par
 \vspace{1.75cm}
\includegraphics[scale=0.4]{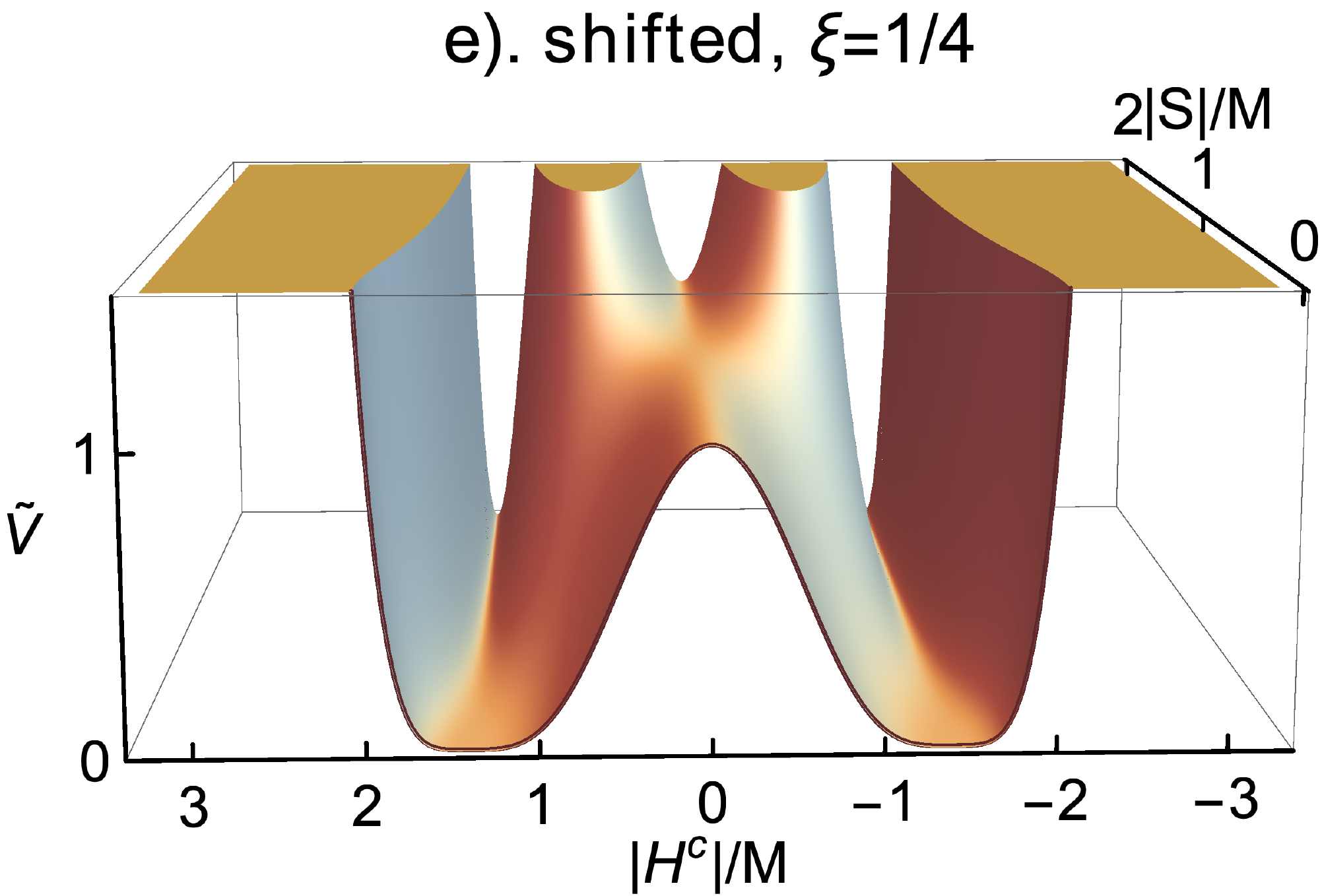}
\includegraphics[scale=0.4]{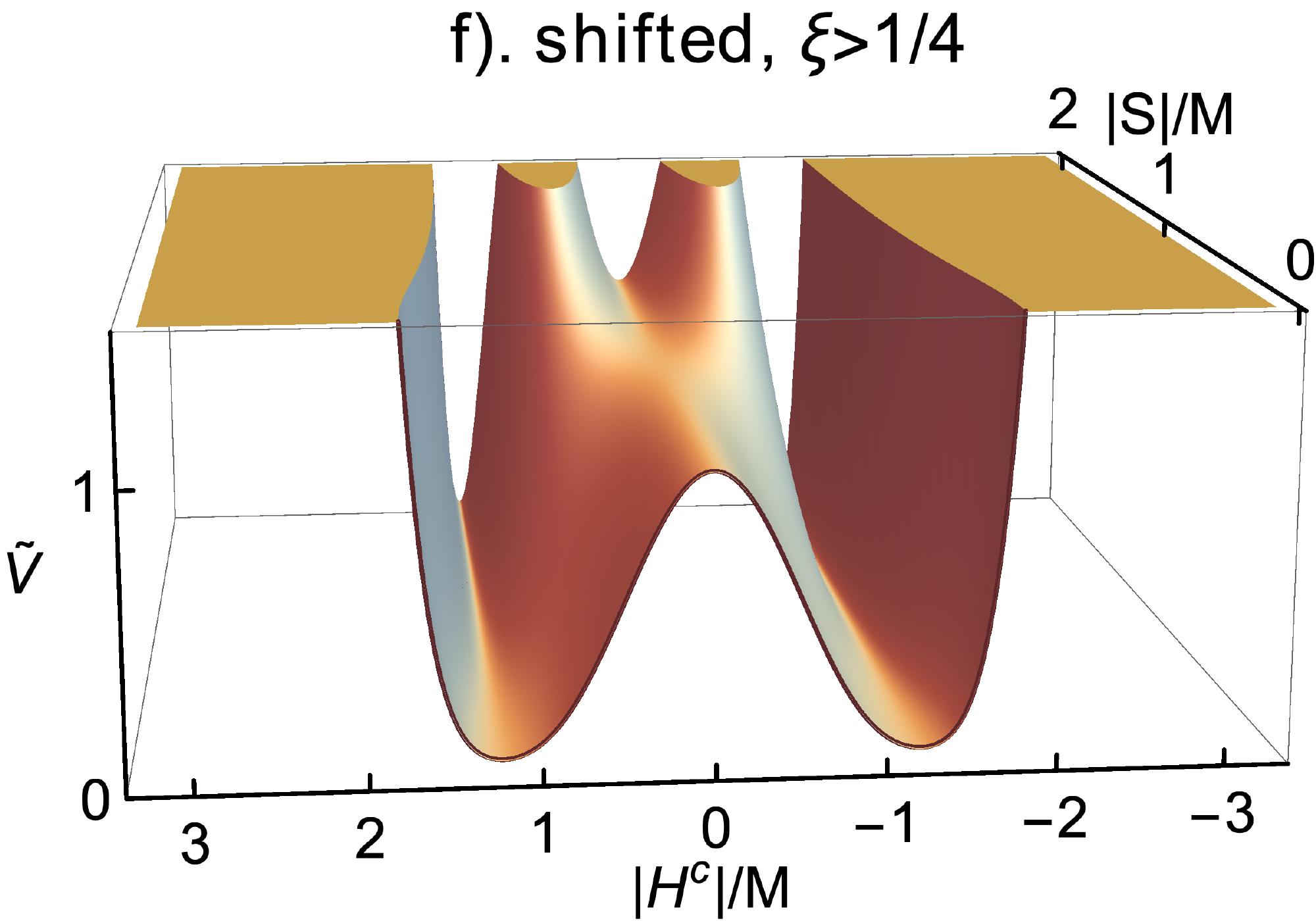}
\caption{\label{pot} The normalized scalar potential $\widetilde V(w,u,z=0)=V/\kappa^2M^4$, where $w=|S|/M$, $u=|H^c|/M$. The standard $\mu$HI case is reproduced in plot (a). Here $u=0$, $w>1$ is the only inflationary valley available in this case and evolves at $w=0$ into a single pair of global SUSY minima with vev $v=\pm M$. For $\xi\neq 0$, in addition to the standard track at $u=u_1$, there are two shifted trajectories at $u=u_2=\pm 1/\sqrt{2\xi}$, for $w>\sqrt{1/8\xi-1/2}$. Plot (b) shows the undesirable situation where the shifted tracks lie higher than the standard track for $\xi<1/8$. 
Plots (c)-(e) are for $\xi=1/8$, $\xi=1/6$, and $\xi=1/4$, respectively. The case $\xi>1/4$ is shown in plot (f), where the minimal $\widetilde{V}$ is nonzero suggesting that the SUSY vacuum corresponds to complex values of the fields. So any feasible choice for $\xi$ lies in the region 
$[1/8, 1/4]$.} 
\end{figure*}

The global SUSY scalar potential obtained from the superpotential in Eq.~(\ref{SP}) is 
\begin{align}
V&=\Big|\kappa\{\overline{H^c} H^c-M^2-\xi \frac {(\overline{ H^c}H^c)^2}{M^2}\}+\lambda h^2\Big|^2+\lambda^2 h^2|S|^2\nonumber \\
&\!+\!\kappa^2|S|^2(|H^c|^2\!+\!|\overline {H^c}|^2)\Big|1\!-\!2\xi \frac{\overline{H^c}H^c}{M^2}\Big|^2\!+\!D\text{-terms,}
\label{potsusy}
\end{align}
where $|h|^2=|h_u|^2+|h_d|^2$. The $D$-flatness requirement implies that $\overline{H^c}=e^{i\theta}
H^{c*}$ and $h_u^i=e^{i\varphi}\epsilon_{ij}h_d^{j*}$, where $\theta$ and $\varphi$ are invariant angles and $\epsilon_{ij}$ is the $2\times 2$ antisymmetric matrix with $\epsilon_{12}=1$. We have proved that, for $h=0$ and $\xi\leq 1/4$, the potential in Eq.~(\ref{potsusy}) is minimized for 
$\theta=0$ in all cases including the shifted inflationary valley -- see below. Therefore, for our purposes here, we can fix $\theta=0$. Moreover, one can show that, on the shifted path, the potential for $h\neq 0$ is minimized at $\varphi=\pi$. Under these circumstances, the scalar potential along the 
$D$-flat direction takes the form:
\begin{align}
V&=\Big|\kappa \big( |H^c|^2-M^2-\xi \frac {|H^c|^4}{M^2} \big)-\lambda |h_d|^2\Big|^2\nonumber \\
&+2\lambda^2 |h_d|^2|S|^2+2\kappa^2|S|^2|H^c|^2\Big|1-2\xi \frac {|H^c|^2}{M^2}\Big|^2,
\end{align}
which on the shifted path is minimized for $h=0$ provided that $\lambda\geq 2\kappa$. This inequality guarantees the stability of the shifted path at $h=0$ and we can safely set $h$ equal to zero from now on. Rotating the complex field $S$ to the real axis by suitable transformations, we can identify the normalized real scalar field $\sigma=\sqrt 2 S$ with the inflaton. Introducing the dimensionless fields
\begin{equation}
w=\frac{|S|}{M},\ \ \   u=\frac{|H^c|}{M},
\end{equation}
the normalized potential $\widetilde V\equiv V/\kappa^2M^4$ takes the form
\begin{equation}
\widetilde V=(u^2-1-\xi u^4)^2 +2w^2u^2(1-2\xi u^2)^2.
\end{equation}
The extrema of the above potential with respect to $u$ are given as:
\begin{subequations}
\begin{align}
u_1&=\ 0\, ,{\label{vac1}}\\
u_2&=\pm\frac{1}{ \sqrt{2\xi }}\,,{\label{vac2}}\\ 
u^\pm_{3}\!&=\!\frac{1}{\sqrt {2\xi}}\sqrt{1\!-\!6 w^2\xi\pm\sqrt{\!-\!4 \xi\!+\!36 \xi ^2 w^4\!-\!8 \xi  w^2\!+\!1}}.{\label{vac3}}
\end{align}
\end{subequations}
These extrema can be visualized with the help of the potential $\widetilde V(u,w)$, plotted in 
Fig.~\ref{pot}, for various values of the parameter $\xi$. 

 In Fig.~\ref{pot}, the standard $\mu$HI case with $\xi=0$ is reproduced in plot (a). In this case, 
$u=0$, $w>1$ is the only inflationary valley available. It evolves at $w=0$ into a single pair of global SUSY minima with vev $v=\pm M$. For $\xi\neq 0$, in addition to the standard track at $u=u_1$, two shifted local minima appear at $u=u_2$ for $w>\sqrt{1/8\xi-1/2}$.
In plot (b) for $\xi<1/8$, the shifted tracks lie higher than the standard track. Following 
Ref.~\cite{Jeannerot:2000sv}, in order to have suitable initial conditions for realizing inflation along the shifted tracks, we assume $\xi\geq 1/8$. The normalized scalar potential $\widetilde V$ is shown in plots (c)-(e) for some realistic values of $\xi$, namely for $\xi=1/8$, $\xi=1/6$, and 
$\xi=1/4$. In the last plot (f) with $\xi>1/4$, we obtain $V_{min}\neq 0$, since the SUSY minimum 
requires complex values of $\overline{H^c}$, $H^c$. So for our analysis, it is appropriate to keep 
$\xi$ within the interval $[1/8, 1/4]$.
 
 As the inflaton slowly rolls down the inflationary valley and enters the waterfall regime at 
$w=\sqrt{1/8\xi-1/2}$, inflation ends due to fast rolling and the system starts oscillating about a vacuum at $w=0$. Note that in the $H^c$ direction there are actually two pairs of vacua at [see Eq.~(\ref{vac3})]
\begin{equation}
 (u_3^\pm)^2 \xrightarrow{w=0} v^2_\pm=\frac{1}{2\xi} [1\pm\sqrt{1-4 \xi}].
 \end{equation}
However, the path leading to $v_-$ appears before the one leading to $v_+$, as explained in Ref.~\cite{Jeannerot:2000sv}.
The necessary slope for realizing inflation in the valley with $w>\sqrt{1/8\xi-1/2}$, $u=u_2$, $z=0$ is generated by the inclusion of the one-loop radiative corrections, the SUGRA corrections, and the soft SUSY breaking terms. The one-loop radiative corrections $V_{loop}$, arising as a result of SUSY breaking on the inflationary path, are calculated using the Coleman-Weinberg formula \cite{Coleman;1973}: 
 \begin{align}\label{CW}
V_{loop}&=\frac{1}{64\pi^2} \sum_i (-1)^{F_i} M_{i}^4\ln\Big(\frac{M_{i}^2(S)}{Q^2}-\frac{3}{2}\Big)\nonumber \\
&=\kappa^2m^4 \left[ \frac{\kappa^2 }{4 \pi ^2}F(x)+\frac{\lambda ^2}{4 \pi ^2} F(y) \right],
  \end{align}
 where $F_i$ and $M_{i}^2$ are the fermion number and squared mass of the \textit{ith} state. The function $F(x)$ is given by
  \begin{align}
F(x)&=\frac{1}{4}[(x^4+1)\ln\big(\frac{x^4-1}{x^4}\Big)+2x^2\ln\Big(\frac{x^2+1}{x^2-1}\Big)\nonumber \\ 
&+2\ln\Big(\frac{2\kappa^2m^2x^2}{Q^2}\Big)-3],
\end{align}
$y=\sqrt{\gamma/2}\ x$ with $\gamma=\lambda/\kappa$, and $x$ is defined in terms of the canonically normalized real inflaton field $\sigma$ as $x=\sigma/m$ with $m^2=M^2(1/4\xi-1)$. 
 The function $F(y)$ exhibits the contribution of the $\mu$ term in the superpotential $W$, and for $\gamma\gtrsim 1$, is expected to play an important role in the predictions of inflationary observables. The renormalization scale $Q$ is set equal to $\sigma_0$, the field value at the pivot scale $k_0=0.05\text{ Mpc}^{-1}$ \cite{Akrami:2018odb}. 

The soft SUSY breaking terms are added in the inflationary potential as:
\begin{equation}
V_{soft}=m_{3/2}\big[z_i\frac{\partial W}{\partial z_i}+(A-3)W +h.c.\big],
\end{equation}
where $A$ is the complex coefficient of the trilinear soft-SUSY-breaking terms.

Trying to reconcile supergravity and cosmic inflation, one runs into the so-called $\eta$ 
problem which arises as the effective inflationary potential is quite 
steep. This leads to large inflaton masses on the order of the Hubble 
parameter $H$ and thus the slow-roll conditions are violated. In hybrid inflationary scenarios, the supergravity corrections can 
easily be brought under control \cite{Lazarides:1996dv,Panagiotakopoulos:1997ej,Dvali:1997uq,Lazarides:1998zf,Dimopoulos:2011ym}. Another potential problem is 
the appearance of anti-de Sitter vacua. However, in hybrid inflation models, these
vacua may be lifted -- for examples see Refs.~\cite{Haba:2005ux,Wu:2016fzp}.

The $F$-term SUGRA scalar potential is evaluated using, 
 \begin{equation}\label{sugra}
  V_{SUGRA}=e^{K/m_{P}^2}(K_{i\bar j}^{-1}D_{z_i}WD_{z_{\bar j}^*} W^*-3m_{P}^{-2} |W|^2),
   \end{equation}  where $z_i\in\{S,\ H^c,\  \overline{H^c},\ h,\ ...\}$ and
  \begin{align}
 K_{i j}& \equiv \frac{\partial^2K}{\partial z_i \partial z^*_{ j}},\nonumber \\
 D_{\, z_{i}}\, W \, &\equiv  \frac{\partial W}{\partial z_i}+ m_{P}^{-2} \frac{\partial K}{ \partial z_i}W,\nonumber \\
 D_{z_{ i}^*}W^*&=(D_{z_i}W)^*. 
\end{align}
The K\"ahler potential $K$ is expanded in inverse powers of $m_P$: 
 \begin{align}\label{fullkahler} 
K&= K_c+\kappa_S  \frac{|S|^4}{4m_P^2}+\kappa_H  \frac{|H^c|^4}{4m_P^2}+\kappa_{\overline H} \frac{|\overline{H^c}|^4}{4m_P^2}+\kappa_h  \frac{|h|^4}{4m_P^2}\nonumber \\ 
&+\kappa_{SH^c}  \frac{|S|^2|H^c|^2}{m_P^2}+\kappa_{S\overline{H^c}}  \frac{|S|^2|\overline{H^c}|^2}{m_P^2}+\kappa_{Sh}  \frac{|S|^2|h|^2}{m_P^2}\nonumber \\
&+\kappa_{H^c\overline{H^c}}  \frac{|H^c|^2|\overline{H^c}|^2}{m_P^2}+\kappa_{H^ch}  \frac{|H^c|^2|h|^2}{m_P^2}+\kappa_{\overline{H^c}h}  \frac{|\overline{H^c}|^2|h|^2}{m_P^2}\nonumber\\
&+\kappa_{SS} \frac{|S|^6}{6m_P^4}+...  ~ ~~,
\end{align}
where the minimal canonical K\"ahler potential $K_c$ is given by
 \begin{equation}\label{mkahler}
 K_c=  |S|^2+|H^c|^2+|\overline{H^c}|^2 +|h^2|.
 \end{equation}

The inflationary potential along the D-flat direction with $|H^c|=|\overline{H^c}|$, stabilized along the $h=0$ direction, and incorporating the SUGRA corrections \cite{Linde:1997sj}, the radiative corrections \cite{Dvali:1994ms}, and the soft-SUSY-breaking terms \cite{Senoguz:2004vu,Rehman:2009nq}, is given by
\begin{align}\label{total}
V(x)&\simeq V_{SUGRA}+ V_{loop}+V_{soft}\nonumber \\
&\simeq\kappa^2m^4\Bigg(\mathcal A+ \frac{1}{2}\mathcal B\Big(\frac{m}{m_P}\Big)^2x^2+\frac{1}{4} \mathcal C\Big(\frac{m }{ m_P}\Big)^4 x^4\nonumber \\
&+ \frac{\kappa^2 }{4 \pi ^2}F(x)+\frac{\lambda ^2}{4 \pi ^2} F(y)\nonumber \\
&+a\frac{m_{3/2}}{\sqrt{2}\kappa m}x+\frac{m_{3/2}^2 }{2 \kappa^2 m^2}x^2+\frac{m_{3/2}^2M^2}{\kappa^2 m^4 \xi }\Bigg).
 \end{align}
Here $\mathcal A$, $\mathcal B$, and $\mathcal C$ are the coefficients of the constant, quadratic, and quartic SUGRA terms, respectively, and are defined in terms of $H_P=(M/m_P)/\sqrt{2\xi}$ as
\begin{equation} \label{coeff}
\mathcal A \!=\!1\!+\!2 c_0 H_P^{2}\!+\!2c_1H_P^{4},\ \ \,
\mathcal B \!=\!-\kappa_S \!+\!2c_2H_P^{2},\ \ \,
\mathcal C \!=\!\frac{\gamma_S}{2},
\end{equation}
where $\gamma_S = 1 + 2 \kappa _S^2-3 \kappa _{SS}-7 \kappa _S/2\ $ \cite{Civiletti:2011qg}. For the inflationary potential along the D-flat and $h=0$ direction, the independently varying parameters 
$c_0$, $c_1$, and $c_2$ for the nonminimal case are the same as the ones 
given in Ref.~\cite{Civiletti:2011qg}. Our choice for these parameters will be shown in the relevant sections. The parameter $a$ depends on $\arg S$ as follows:
\begin{equation}
a = 2\left| 2-A+\frac{A}{2\xi}\right| \cos [\arg S+\arg (2-A+\frac{A}{2\xi})].
\end{equation}
Assuming negligible variation in $\arg S$, with $a=-1$, the scalar spectral index $n_s$ is expected to lie within the experimental range \cite{Rehman:2009nq, Civiletti:2011qg}. This could also be achieved by taking an intermediate-scale, negative soft mass-squared term for the inflaton \cite{Rehman:2009yj}. But with the nonminimal terms in the  K\"ahler potential, one can also obtain the central value of $n_s$ with TeV-scale soft masses even for $a=1$ \cite{BasteroGil:2006cm, urRehman:2006hu}.  
The variation in $\arg S$ with general initial condition has been studied in Refs.~\cite{Senoguz:2004vu, urRehman:2006hu, Buchmuller:2014epa}. 

The slow-roll parameters are defined by
 \begin{equation}
 \epsilon\!=\!\frac{m_{p}^2}{2m^2}\Big(\frac{V'}{V}\Big)^2\!,\ \  \eta\!=\!\frac{m_{p}^2}{m^2}\Big(\frac{V''}{V}\Big),\ \   \zeta^2\!=\!\frac{m_{p}^4}{m^4}\Big(\frac{V'V'''}{V^2}\Big),
  \end{equation}
where the primes denote derivatives with respect to $x$. The scalar spectral index $n_s$, the tensor-to-scalar ratio $r$, the running of the scalar spectral index $dn_s/d\ln k$, and the scalar power spectrum amplitude $A_s$, to leading order in the slow-roll approximation, are as follows:
\begin{subequations}
\label{para}
  \begin{align}
  n_s&\simeq 1-6\epsilon+2\eta,\\ 
    r&\simeq 16\epsilon,\\
   \frac{dn_s}{d\ln k}&\simeq 16\epsilon\eta-24\epsilon^2-2\zeta^2,\\ 
     A_{s}(k_0)&=\frac{1}{12\pi^2}\Big(\frac{m}{ m_P}\Big)^2\Big|\frac{V^3/{V'}^{2}}{m_P^4}\Big|_{x_0},\label{As}
  \end{align}
  \end{subequations}
 where $A_s(k_0)=2.196\times10^{-9}$ and $x_0$ denotes the value of $x$ at the pivot scale $k_0=0.05\text{ Mpc}^{-1}$ \cite{Akrami:2018odb}. For the numerical estimation of the inflationary predictions, these relations are used up to second order in the slow-roll parameters. 
 
Assuming a standard thermal history, the number of $e$-folds $N_0$ between the horizon exit of the pivot scale and the end of inflation is 
   \begin{align}\label{n0}
   N_0&=\Big(\frac{m}{m_P}\Big)^2\int_{1}^{x_0}\Big(\frac{V}{V'}\Big)dx\\ 
   &=53+\frac{1}{3}\ln\Big(\frac{T_r}{10^9\text{ GeV}}\Big)+\frac{2}{3}\ln\Big(\frac{\sqrt\kappa m}{10^{15}\text{ GeV}}\Big).\nonumber
   \end{align}
 The reheat temperature $T_r$ is approximated by
    \begin{equation}\label{reheat}
    T_r\approx\sqrt[4]{\frac{90}{\pi^2{g_*}}} \sqrt{{\Gamma_S } {m_P}},\\
    \end{equation}
   where $g_*=228.75$ for MSSM and $\Gamma_S$ is the inflaton decay width. 
  From the $\mu$-term coupling $\lambda Sh^2$ in Eq.~(\ref{SP}), we see that the inflaton can decay into a pair of Higgsinos $\widetilde h_u$, $\widetilde h_d$ with a decay width
  \begin{equation}\label{gamma}
 \Gamma_S(S \rightarrow \widetilde h_u\widetilde h_d)=\frac{\lambda ^2 }{8 \pi }m_{\text{infl}},
   \end{equation}
 where 
  \begin{equation}
  m_{\text{infl}} =\sqrt{2}\kappa v\Big(1-\frac{2\xi v^2}{M^2}\Big)=2\kappa m \sqrt{1-\sqrt{1-4 \xi}}
	\label{minfl}
  \end{equation}
is the inflaton mass \cite{Jeannerot:2000sv}.
 The reheat temperature, the inflaton decay width, and the inflaton mass defined above in  
Eqs.~(\ref{reheat})-(\ref{minfl}) are used together with Eq.~(\ref{n0}) in order to derive the numerical predictions for the present inflationary scenario.
\section{\label{min}$\bm{\mu}$-hybrid inflation with minimal K\"ahler potential}
 The inflationary potential corresponding to the minimal K\"ahler potential $K_c$ in 
Eq.~(\ref{mkahler}) is easily transcribed from Eq.~(\ref{total}) as follows:  
 \begin{align} \label{potential}
V(x)&\simeq \kappa^2m^4\Bigg(1+ 2\Big(\frac{M}{\sqrt{2\xi} m_P}\Big)^2+2\Big(\frac{M}{\sqrt{2\xi} m_P}\Big)^4\nonumber \\
&+ \Big(\frac{M}{\sqrt{2\xi} m_P}\Big)^2 \Big(\frac{m}{m_P}\Big)^2x^2+\frac{1}{8}\Big(\frac{m }{ m_P}\Big)^4 x^4\nonumber \\
&+ \frac{\kappa^2 }{4 \pi ^2}F(x)+\frac{\lambda ^2}{4 \pi ^2} F(y)\nonumber \\
&+a\frac{m_{3/2}}{\sqrt{2}\kappa m}x+\frac{m_{3/2}^2 }{2 k^2 m^2}x^2+\frac{m_{3/2}^2M^2}{k^2 m^4 \xi }\Bigg),
 \end{align}
since, in this case, $\mathcal C =1/2$, $c_0=c_1=c_2=1$ and, thus, the coefficients $\mathcal A =1+2  (H_P^{2}+H_P^{4})$, $\mathcal B =2H_P^{2}$.
 \begin{figure}
\includegraphics[scale=0.425]{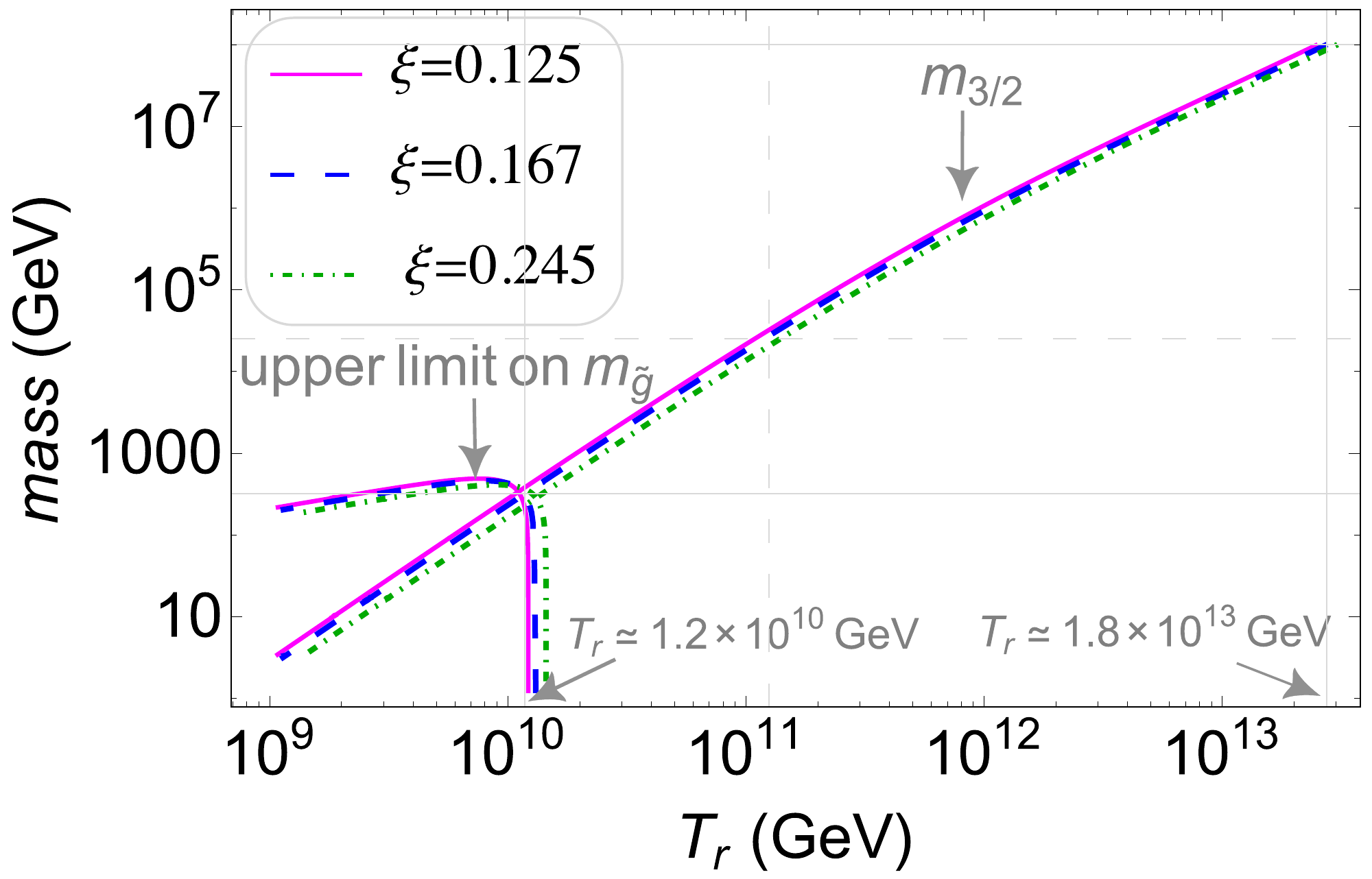}
\caption{\label{fig:M3mgTr}Plot of the gravitino mass $m_{3/2}$ versus the reheat temperature $T_r$ for successful inflation, and of the upper limit on the gluino mass $m_{\tilde g}$ assuming a stable gravitino LSP. The solid-magenta, dashed-blue, dot-dashed-green curves correspond to 
$\xi = 0.125,\ 0.167,\ 0.245$ respectively for the minimal K\"{a}hler potential with the conditions 
$n_s\simeq 0.964$, $A_s(k_0)=2.196\times10^{-9}$, $\gamma=2$, and $a=-1$. The intersection point where 
$m_{3/2}$ coincides with the upper limit on $m_{\tilde g}$, for the central value of $\xi$, is at 
$T_r\simeq 1.2\times 10^{10}$~GeV and $m_{3/2}\simeq 325$~GeV. The maximum value of the gluino mass in the region where $m_{3/2}$ is smaller than the upper limit on $m_{\tilde g}$ is $m_{\tilde g}\sim 500$~GeV, which is lower than the lower LHC bound on the gluino mass ($m_{\tilde{g}}\gtrsim 
1$~TeV). Hence, the gravitino LSP scenario is inconsistent. For the unstable gravitino scenario, 
$m_{3/2}\simeq 25{\rm\ TeV}$ corresponds to $T_r\sim 10^{11}{\rm\ GeV}$ as shown by the vertical dashed-gray line.} 
\end{figure} 

 In Fig.~\ref{fig:M3mgTr}, we plot the gravitino mass $m_{3/2}$ versus the reheat temperature $T_r$ as constrained by inflation. The solid-magenta, dashed-blue, dot-dashed-green curves correspond to $\xi = 
0.125,\ 0.167,\ 0.245$ respectively for the minimal K\"{a}hler potential with the conditions $n_s\simeq 0.964$, $A_s(k_0)=2.196\times10^{-9}$, $\gamma=2$, and $a=-1$. The lower bound on the reheat temperature $T_r\gtrsim 10^{9}$ GeV is obtained for a gravitino mass $m_{3/2} \gtrsim 3.5{\rm\ GeV}$ with a $0.1\%$ fine-tuning of the difference $x_0 -1$. 

Following the same line of argument as in Refs.~\cite{Okada:2015vka,Rehman:2017gkm}, the shifted 
$\mu$HI with minimal $K$ is analyzed for the following three cases: 
\begin{enumerate}
\item{stable gravitino LSP}; 
\item{unstable long-lived gravitino with $m_{3/2}<25{\rm\ TeV}$}; 
\item{unstable short-lived gravitino with $m_{3/2}>25{\rm\ TeV}$.} 
\end{enumerate}

The relic gravitino abundance, in the case of a stable gravitino LSP, is given \cite{Bolz:2000fu} by
\begin{equation}\label{gluino}
\Omega_\text{3/2}h^2=0.08 \Big(\frac{T_r}{10^{10}\text{GeV}}\Big)\Big(\frac{m_\text{3/2}}{1\text{ TeV}}\Big)\Big(1+\frac{m_{\tilde g}^2}{3m_\text{3/2}^2}\Big),
\end{equation}
where $m_{\tilde g}$ is the gluino mass.
We require that $\Omega_\text{\tiny{3/2}}h^2$ does not exceed the observed DM relic abundance, that is $\Omega_\text{\tiny{3/2}}h^2\lesssim 0.12$ \cite{Akrami:2018odb}. Using
Eq.~(\ref{gluino}), we then plot in Fig.~\ref{fig:M3mgTr} the resulting upper limit on the gluino mass 
$m_{\tilde g}$. The point where $m_{3/2}$ and the upper bound on $m_{\tilde g}$ coincide, for the central value of $\xi$ (i.e. $\xi=0.167$), lies at $T_r\simeq 1.2\times10^{10}$~GeV and $m_{3/2}\simeq 325$~GeV as shown by the intersection of the corresponding curves. Our assumption for a gravitino LSP holds for $T_r$ values below this intersection point, that is for $T_r \lesssim 1.2\times10^{10}$~GeV, 
$m_{3/2} \lesssim 325$~GeV. However, the maximum value of the gluino mass in this region is 
$m_{\tilde g}\sim 500$~GeV which is lower than the lower bound on the gluino mass $m_{\tilde{g}}\gtrsim 1$~TeV from the search for supersymmetry at the LHC \cite{Tanabashi:2018oca}. Consequently, we run into inconsistency and the case of a stable gravitino LSP with a minimal K\"ahler potential is ruled out.

In the second case, the long-lived unstable gravitino will decay after big bang nucleosynthesis (BBN), and so one has to take into account the BBN bounds on the reheat temperature which are the following \cite{Khlopov:1993ye, Kawasaki:2004qu, Kawasaki:2017bqm}: 
 \begin{align}\label{bbn}
 T_r&\lesssim 3\times(10^5-10^6) \text{ GeV},  &m_{3/2}&\sim 1\text{ TeV},\nonumber \\
T_r&\lesssim 2\times10^9 \text{ GeV},  &m_{3/2}&\sim 10 \text{ TeV}.
 \end{align} 
 The bounds on the reheat temperature from the inflationary constraints for gravitino masses $1{\rm\ and}\ 10{\rm\ TeV}$  are $T_r\gtrsim 2.2\times10^{10}{\rm\ GeV}$ and $7.5\times10^{10}{\rm\ GeV}$ respectively (see Fig.~\ref{fig:M3mgTr}). These are clearly inconsistent with the above mentioned BBN bounds, and so the unstable long-lived gravitino scenario is not viable. 
 
 Lastly, for the unstable short-lived gravitino case, we compute the LSP lightest neutralino ($\tilde \chi_1^0$) density produced by the gravitino decay and constrain it to be smaller than the observed DM relic density. For reheat temperature $T_r\gtrsim 10^{11}{\rm\ GeV}$ with $m_{3/2}>25{\rm\ TeV}$ (see Fig.~\ref{fig:M3mgTr}), the resulting bound on the neutralino mass $m_{\tilde\chi_1^0}$ comes out to be inconsistent with the lower limit set on this mass $m_{\tilde\chi_1^0}\gtrsim18{\rm\ GeV}$ in 
Ref.~\cite{Hooper:2002nq}. To circumvent this, the LSP neutralino is assumed to be in thermal equilibrium during gravitino decay, whereby the neutralino abundance is independent of the gravitino yield. For an  unstable gravitino, the lifetime is (see Fig.~1 of Ref.~\cite{Kawasaki:2008qe})
 \begin{equation}\label{tau}
 \tau_{3/2}\simeq 1.6\times 10^4\Big(\frac{1\ \text{TeV}}{m_{3/2}}\Big)^3{\rm sec}.\\
 \end{equation}
Now for a typical value of the neutralino freeze-out temperature, $T_F\simeq 0.05\  
m_{\tilde\chi^0_1}$, the gravitino lifetime is estimated to be
 \begin{equation}\label{tau2}
 \tau_{3/2}\lesssim 10^{-11}\Big(\frac{1\ \text{TeV}}{m_{\tilde\chi^0_1}}\Big)^2\text{sec}.\\
 \end{equation}
Comparing Eq.~(\ref{tau}) and Eq.~(\ref{tau2}), we obtain a bound on $m_{3/2}$,
 \begin{equation}
 m_{3/2} \gtrsim 10^8 \Big(\frac {m_{\tilde\chi_1^0}}{2\text{ TeV}}\Big)^{2/3}\text{GeV}.
 \end{equation} 
Thus, minimal shifted $\mu$HI conforms with the conclusion of the standard case
\cite{Okada:2015vka, Rehman:2017gkm} by requiring split-SUSY with an intermediate-scale gravitino mass and reheat temperature $ T_r\gtrsim 1.8 \times10^{13}{\rm\ GeV}$ (see Fig.~\ref{fig:M3mgTr}). To check whether the shifted $\mu$HI scenario is also compatible with low reheat temperature (i.e. $T_r\lesssim 10^{12}-10^{8}~{\rm\ GeV}$ \cite{Khlopov:1984pf})
and TeV-scale soft SUSY breaking, we employ nonminimal K\"ahler potential in the next section. 
\section{\label{NM}$\bm{\mu}$-hybrid inflation with nonminimal K\"{a}hler potential }
The nonminimal K\"ahler potential used in the following analysis is
 \begin{equation}\label{nonminKahler} 
K=  K_c+\kappa_S  \frac{|S|^4}{4m_P^2}+\kappa_{SS} \frac{|S|^6}{6m_P^4},
\end{equation}
which includes only the nonminimal couplings of interest $\kappa_S$ and $\kappa_{SS}$. (For a somewhat different approach to $\mu$-hybrid inflation with nonminimal $K$, see Ref.~\cite{Okada:2017rbf}). 
Thus, for the nonminimal scenario we take $c_0=c_1=1$ and $c_2=1-\kappa_S$ in Eq.~(\ref{coeff}) \cite{Civiletti:2011qg}.
Using these values the potential of the system can easily be read off from Eq.~(\ref{total}). 

It is worth noting that with the nonminimal K\"ahler potential we can realize the central value of $n_s$ with TeV-scale soft masses even for $a=1$ \cite{BasteroGil:2006cm, urRehman:2006hu}. Our study is conducted in two parts, described separately in the following subsections, first with $\kappa_{SS}=0$ and then by allowing $\kappa_{SS}$ to be nonzero. The appearance of a negative mass term with a single nonminimal coupling $\kappa_S$ in the potential in Eq.~(\ref{total}) is expected to lead to red-tilted inflation with low reheat temperature, as for standard $\mu$HI (see Ref.~\cite{Rehman:2017gkm}). Furthermore for nonzero $\kappa_{SS}$, the possible larger $r$ solutions leading to observable gravity waves are also anticipated. 
These expectations along with the impact of an additional parameter $\xi$ on inflationary predictions are discussed below.
 \subsection{Low reheat temperature and the gravitino problem}\label{nm}
 \begin{figure}
\includegraphics[scale=0.42]{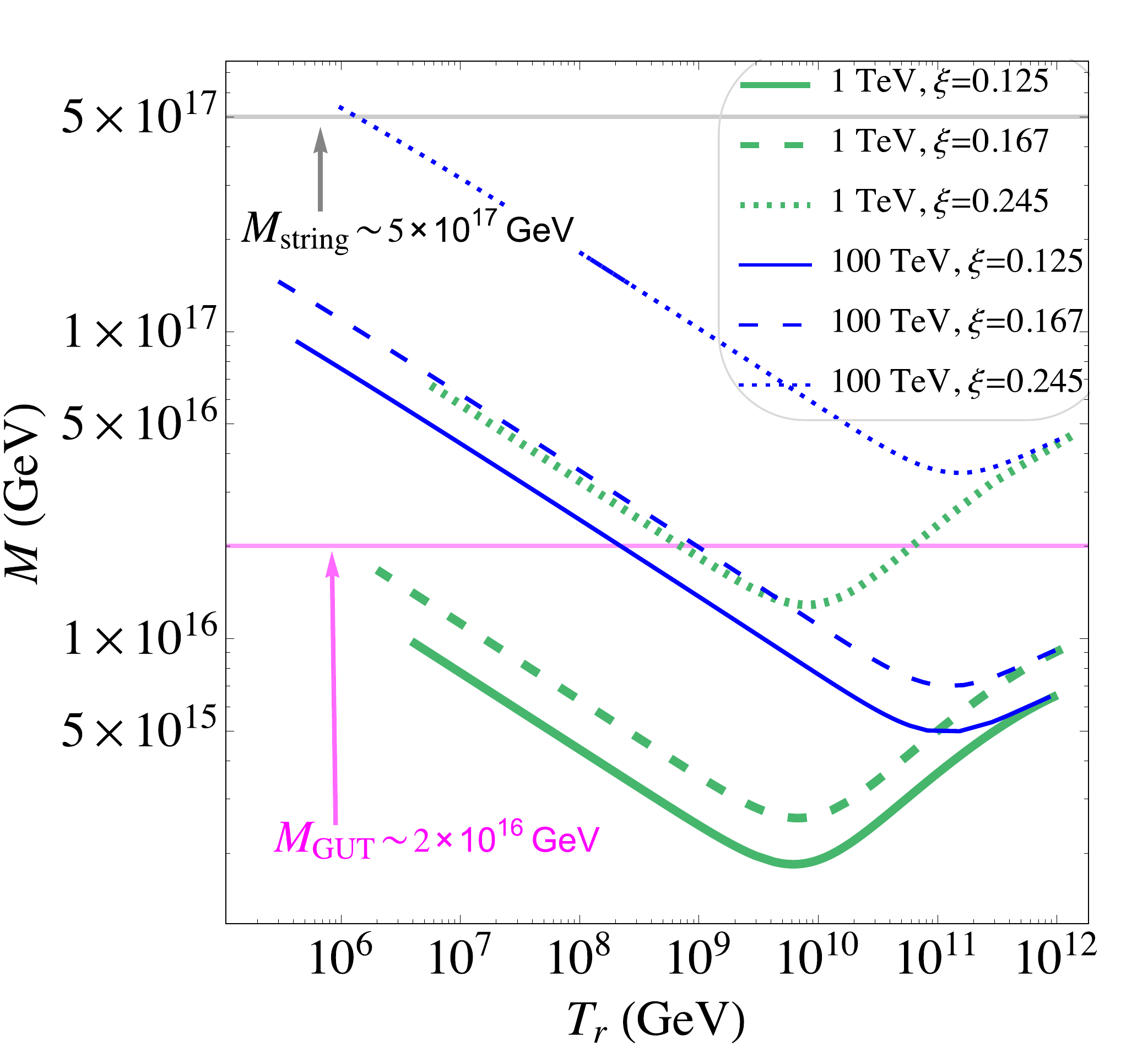}
\includegraphics[scale=0.395]{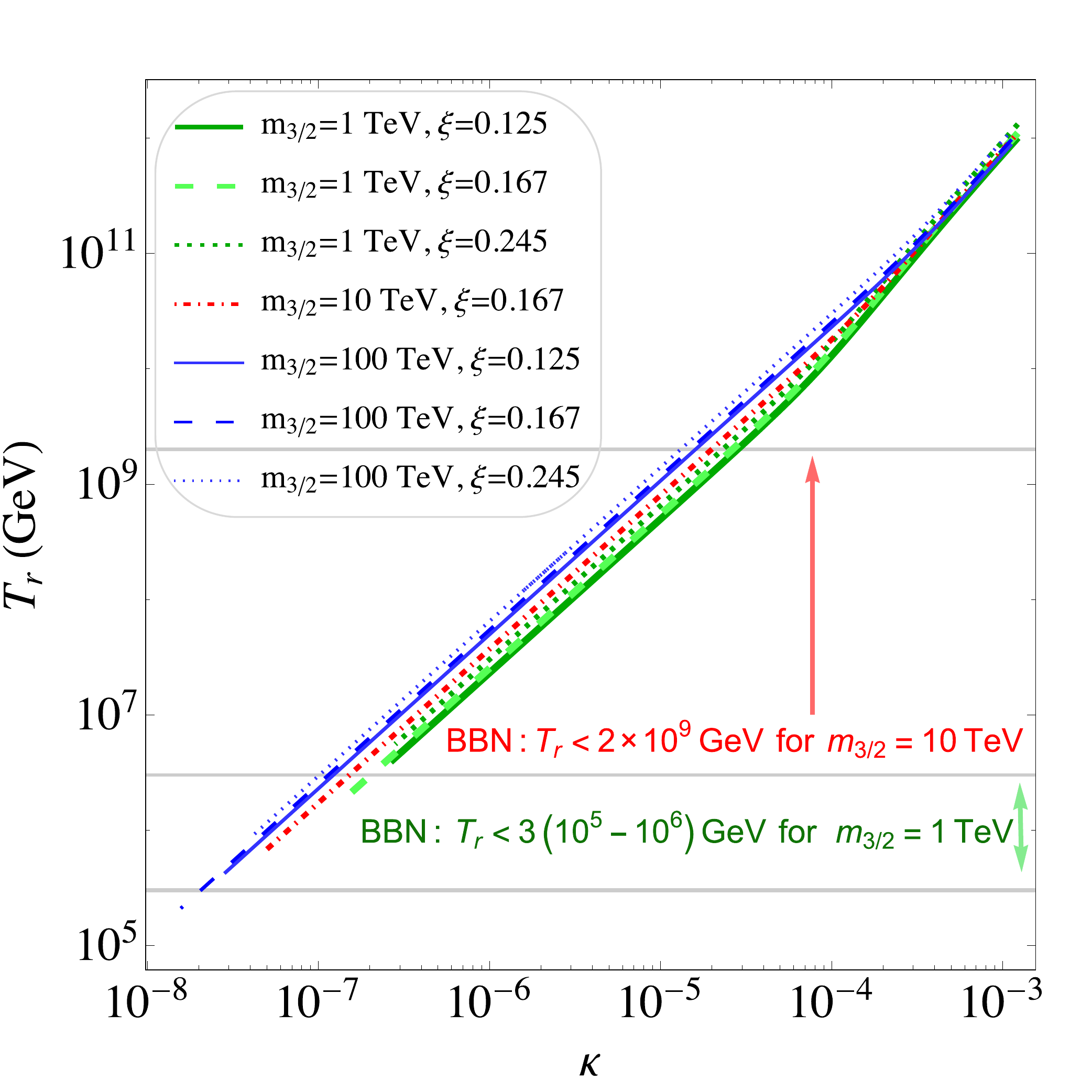}
\caption{\label{MT} The mass scale $M$ versus the reheat temperature $T_r$ and $T_r$ versus $\kappa$, for gravitino mass equal to $1{\rm\ TeV}$ (thick-green curves), 10~TeV (dot-dashed-red curve), and 100~TeV (thin-blue curves). The scalar spectral index $n_s=0.9655$, $\kappa_S=0.02$, $\kappa_{SS}=0$, and $\gamma=2$. The solid, dashed, and dotted curves are for $\xi=0.125$, $0.167$, and $0.245$ respectively.} 
\end{figure}
  \begin{figure}
 \includegraphics[scale=0.425]{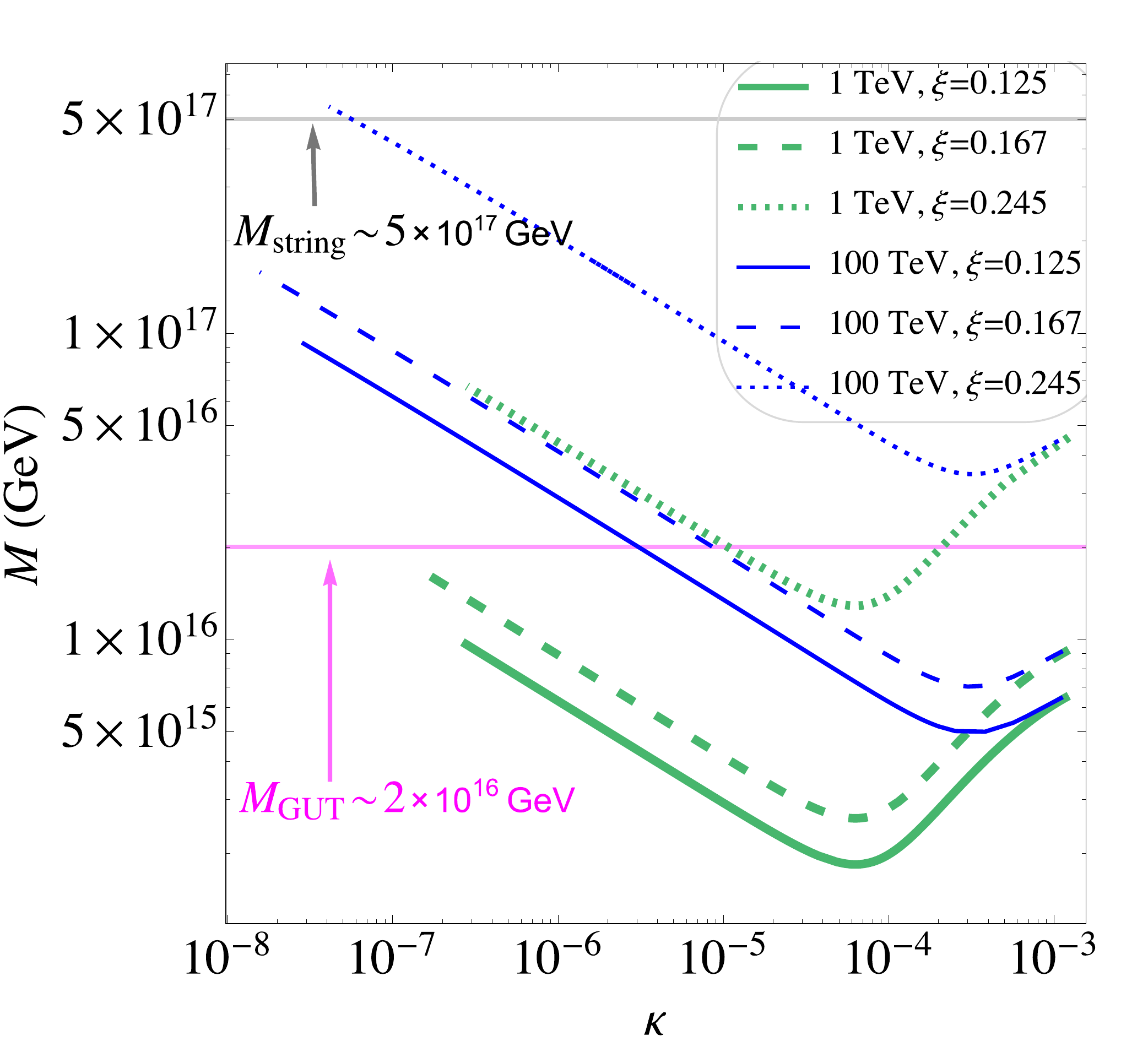}
  \includegraphics[scale=0.425]{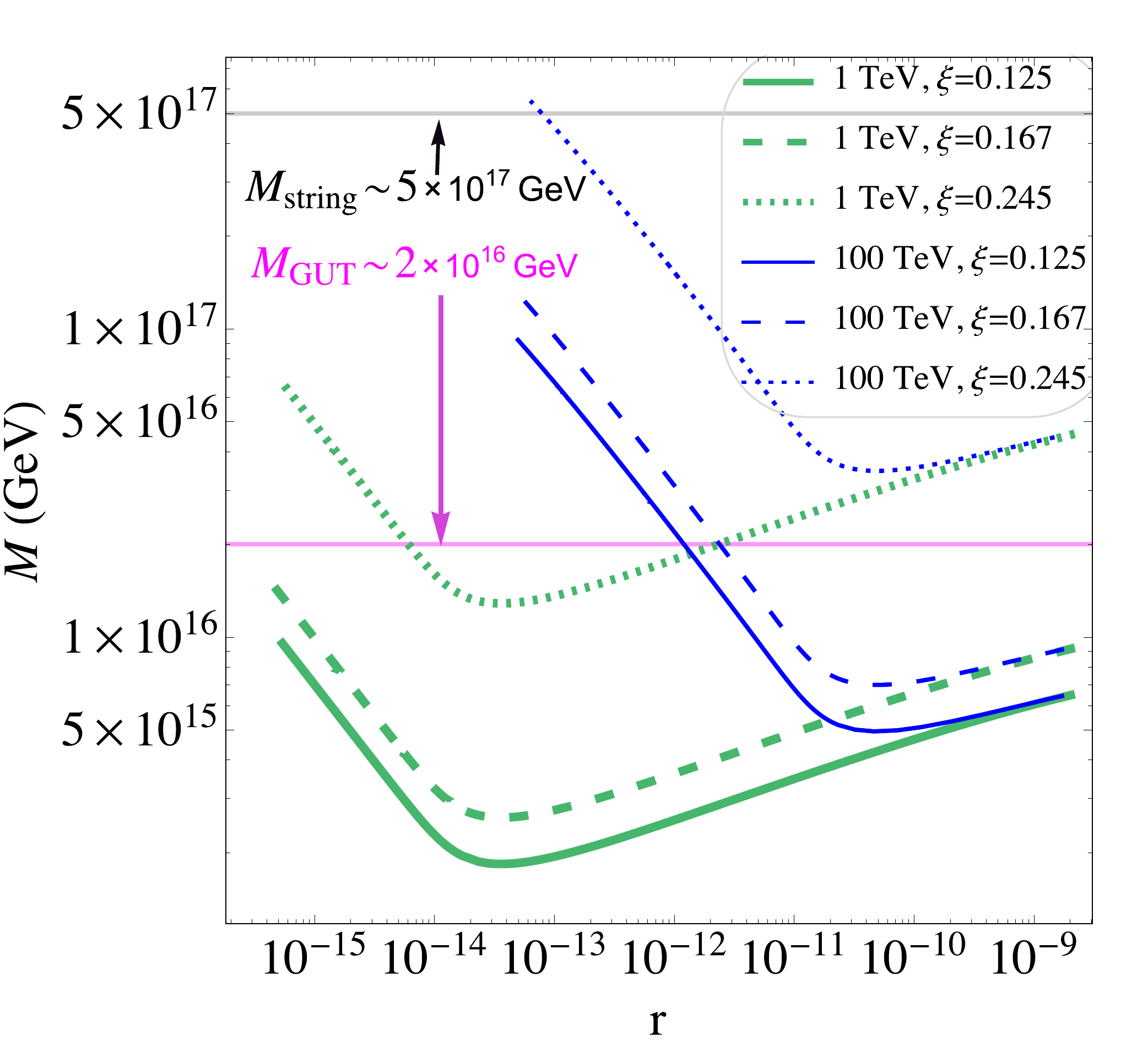}
\caption{\label{mrk} The mass scale $M$ versus $\kappa$ and the tensor-to-scalar ratio 
$r$, for gravitino mass equal to $1{\rm\ TeV}$ (thick-green curves) and 100~TeV (thin-blue curves). We fix the scalar spectral index $n_s=0.9655$, $\kappa_S=0.02$, $\kappa_{SS}=0$, and $\gamma=2$. We consider three values of $\xi$, namely $\xi=0.125,\ 0.167,\ {\rm and}\ 0.245$ corresponding to the solid, dashed, and dotted curves respectively.} 
\end{figure}
 \begin{figure}
  \includegraphics[scale=0.425]{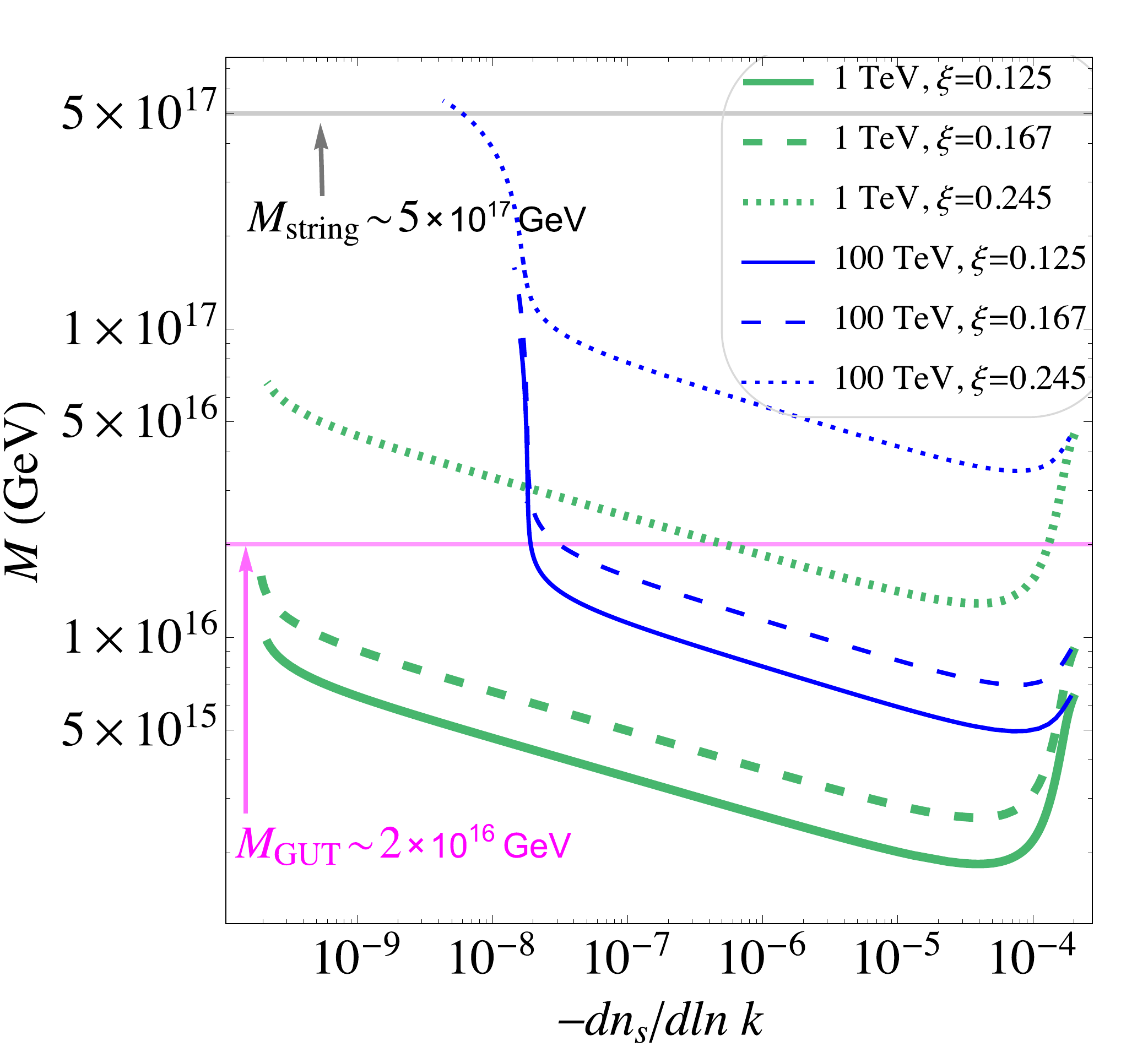}
\includegraphics[scale=0.425]{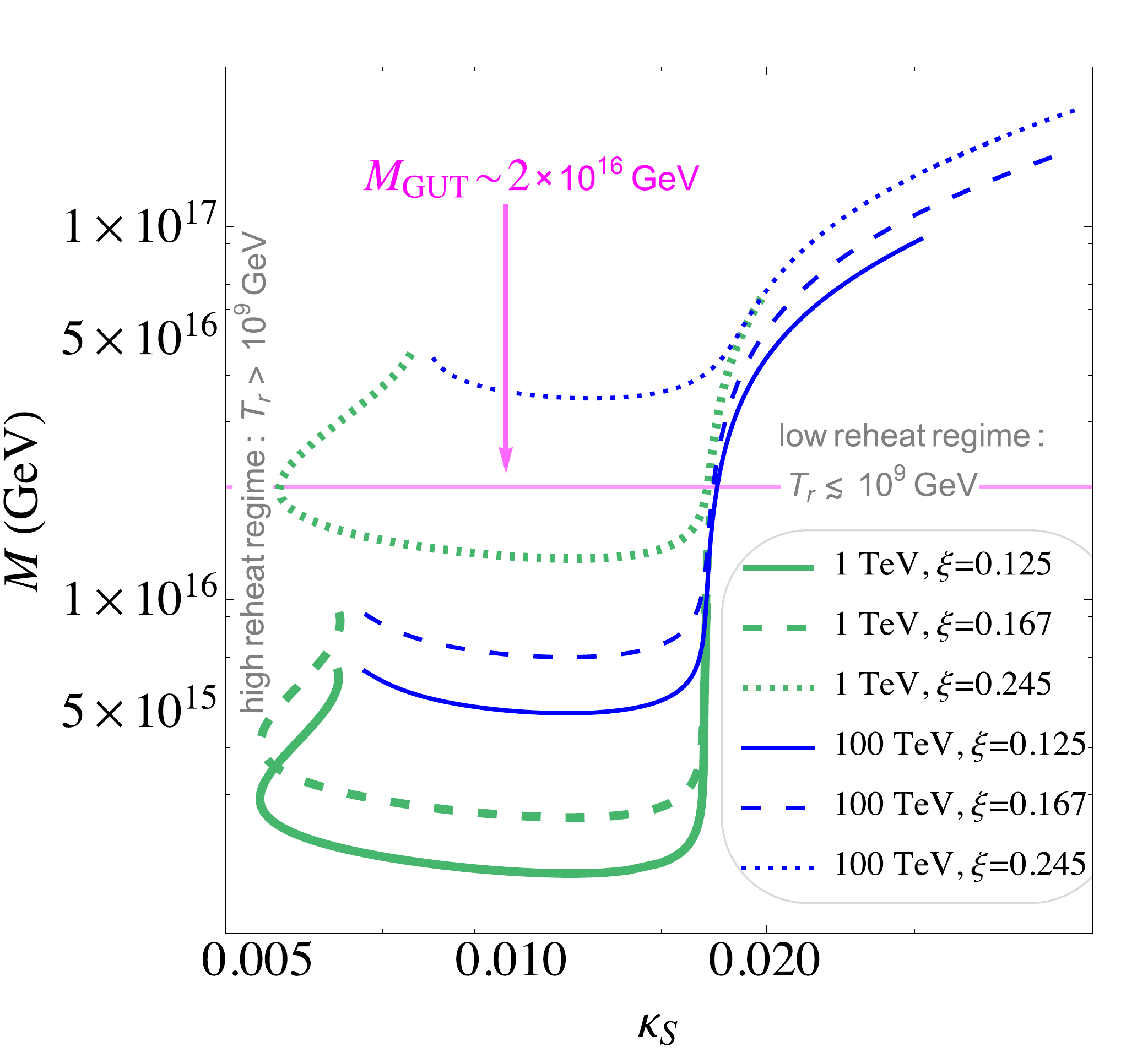}
\caption{\label{mrun}The mass scale $M$ versus the running of spectral index $-dn_s/d\ln k$ and $\kappa_S$, for gravitino mass of $1{\rm\ TeV}$ (thick-green curves) and 100~TeV (thin-blue curves). We fix the scalar spectral index $n_s=0.9655$, $\kappa_S=0.02$,  $\kappa_{SS}=0$, and $\gamma=2$. The parameter $\xi=0.125,\ 0.167,\ {\rm and}\ 0.245$  corresponding to the solid, dashed, and dotted curves respectively.} 
\end{figure}
 Incorporating the inflationary constraints and the nonminimal $K$ in Eq.~(\ref{nonminKahler}) with $\kappa_{SS}=0$, we summarize some of the results depicting the main features of nonminimal shifted $\mu$HI in Figs.~\ref{MT}\,--\,\ref{mrun}. From these figures it is clear that with low reheat temperature we can obtain a higher mass scale $M$ ranging from $5\times 10^{15}{\rm\ GeV}$ to the string scale $5\times 10^{17}{\rm\ GeV}$. The reheat temperature is lowered by nearly half an order of magnitude in the shifted $\mu$HI as compared to the standard $\mu$HI (see Fig.~2 of Ref.~\cite{Rehman:2017gkm}), as can be seen from Fig.~\ref{MT}. Also, it is not surprising that around $\kappa \sim 10^{-3}$ the system is oblivious to the gravitino mass, since the contribution of the linear term becomes less important compared with the SUGRA or radiative corrections \cite{BasteroGil:2006cm}. The interesting new feature is due to the presence of another parameter $\xi\,$, whose effect is to increase the range of mass scale $M$. For a particular value of $\kappa$, say $\kappa \sim 10^{-6}$, and $m_{3/2}=1{\rm\ TeV}$, a wider range of $M\simeq 5\times(10^{15}-10^{16}){\rm\ GeV}$ exists, corresponding to $\xi$ in the range $0.125\leq\xi\leq 0.245$ (see Fig.~\ref{mrk}). So there is an order of magnitude increase in the spread of $M$, compared with standard $\mu$HI, where the maximum value is $M\sim 8\times 10^{15}{\rm\ GeV}$ corresponding to the lowest reheat temperature $T_r\sim 6 \times 10^6{\rm\ GeV}$, with gravitino of mass $1{\rm\ TeV}$ \cite{Rehman:2017gkm}. This maximum value has now increased to $M\simeq(9\times10^{15}-7\times 10^{16}){\rm\ GeV}$ with $\xi$ in the range $0.125\leq\xi\leq 0.245$. Also, the lower plot of Fig.~\ref{mrk}  shows the variation of $M$ with respect to the tensor-to-scalar ratio $r$ with $r\lesssim 10^{-9}$, which is experimentally inaccessible in the foreseeable future 
\cite{Andre:2013afa,Matsumura:2013aja,Kogut:2011,Finelli:2016cyd}. 

As Fig.~\ref{mrun} shows, the running of the scalar spectral index $dn_s/d\ln k$ also turns out to be small in the present scenario, namely $10^{-10}\lesssim -dn_s/d\ln k\lesssim 10^{-4}$, which is a common feature of small field models. The nonminimal K\"{a}hler coupling $\kappa_S$ remains constant in the low reheat temperature range as can be seen from the lower plot of Fig.~\ref{mrun}, since the radiative and the quartic-SUGRA corrections can be neglected in this regime. The scalar spectral index $n_s$ in the low reheat temperature region is $n_s\simeq 1-2\kappa_S$ \cite{King:1997ia}, and so for the central value of the scalar spectral index $n_s= 0.9655$, one obtains $\kappa_S=0.0173$, as exemplified by Fig.~\ref{mrun}. To explore larger values of $r$, we will make use of the freedom provided by the second nonrenormalizable coupling $\kappa_{SS}$ in the next section. Note that the number of
e-folds $N_0$ in Eq.~(\ref{n0}) generally ranges between about 47 and 56.

 Proceeding next to the role of the gravitino in cosmology, one can read off the lower bounds on the reheat temperature $T_r$ from Fig.~\ref{MT}. Since, at low reheat temperatures, inflation occurs near the waterfall region (with $x_0$ close to 1), we devised a criterion by allowing only $0.01\%$ fine-tuning on the difference $x_0 -1$. This yields
\begin{equation}\label{tuning}
T_r\! \gtrsim\! 2\times 10^6\!,7\times10^5\!,\, 2\times10^5\,\text{GeV for } 
m_{3/2}\!=\!1,\, 10,\,100\,\text{TeV}.
\end{equation}
For the first scenario with the gravitino being the LSP in shifted $\mu$HI with nonminimal K\"ahler potential, the upper bounds on the reheat temperature obtained in Ref.~\cite{Rehman:2017gkm} (see Fig.~3 and Eq.~(30) in this reference) are $T_r\lesssim 2 \times(10^{10},\  10^9,\  10^8)\  \text{GeV\ for }m_{3/2}=1,\ 10,\ 100\  \text{TeV}$ respectively. These upper bounds on $T_r$ are consistent with the lower bounds in Eq.~(\ref{tuning}), and so the scenario with the gravitino as LSP can be consistently realized in the nonminimal K\"{a}hler case.

 For the second possibility, namely an unstable long-lived gravitino (with $m_{3/2}\lesssim 25\  {\rm TeV}$), comparison of Eqs.~(\ref{bbn}) and Eq.~(\ref{tuning}) reveals that an $1 {\rm\ TeV}$ gravitino is marginally ruled out but a $10\  {\rm TeV}$ gravitino lies comfortably within the BBN bounds.

 For the third scenario of a short-lived gravitino (for instance with mass $m_{3/2}=100\  {\rm TeV}$), the gravitino decays before BBN, and so the BBN bounds on the reheat temperature no longer apply. The gravitino decays into the LSP neutralino $\tilde \chi_1^0\, $. We find that the resulting neutralino abundance is given by
 \begin{equation}\label{relic}
 \Omega_{\tilde \chi_1^0}h^2\simeq 2.8\times10^{11}\times Y_{3/2} \Big(\frac{m_{\tilde \chi_1^0}}{1\text{ TeV}}\Big),
 \end{equation}
 where the gravitino yield 
  \begin{equation}\label{yield}
 Y_{3/2}\simeq2.3\times 10^{-12}\Big(\frac{T_r}{10^{10}\text{ GeV}}\Big)
 \end{equation}
 is acceptable over the range $T_r\sim 10^5\, \text{GeV}-10^{12}\  {\rm GeV}$\,
\cite{Kawasaki:2008qe}.
The LSP (lightest neutralino) density produced by the gravitino decay should not exceed the observed DM relic density $ \Omega^\text{\tiny{obs}}_\text{\tiny{DM}}h^2\simeq 0.12$ \cite{Akrami:2018odb}. The resulting bound on the lightest neutralino mass 
\begin{equation}
  m_{\tilde\chi_1^0} \lesssim (18-10^6)\  \text{GeV for } 10^{11}\  \text{GeV}\gtrsim T_r\gtrsim 2\times10^5\  \text{GeV}
 \end{equation}
turns out to be less restrictive than the corresponding bound from the abundance of the lightest neutralino from the gravitino decay in the case of standard $\mu$HI. Indeed, the non-LSP gravitino 
with $m_{3/2} \sim 100{\rm\ TeV}$ is acceptable in a larger domain, namely $10^{5}\  \text{GeV} 
\lesssim T_r \lesssim10^{11}{\rm\ GeV}$. There is nearly an order of magnitude decrease in the acceptable lower reheat temperature as compared with the standard $\mu$HI.
Note that the lower limit on the neutralino mass,  $m_{\tilde\chi_1^0}\gtrsim18{\rm\ GeV}$, is obtained in Ref.~\cite{Hooper:2002nq} by employing a minimal set of theoretical assumptions.  In conclusion the shifted $\mu$HI is successful with $m_{3/2}\sim1-100{\rm\ TeV}$ and low reheat temperatures.
  \subsection{Large $r$ solutions or observable gravity waves}
    \begin{figure}[ht!]
\includegraphics[scale=0.425]{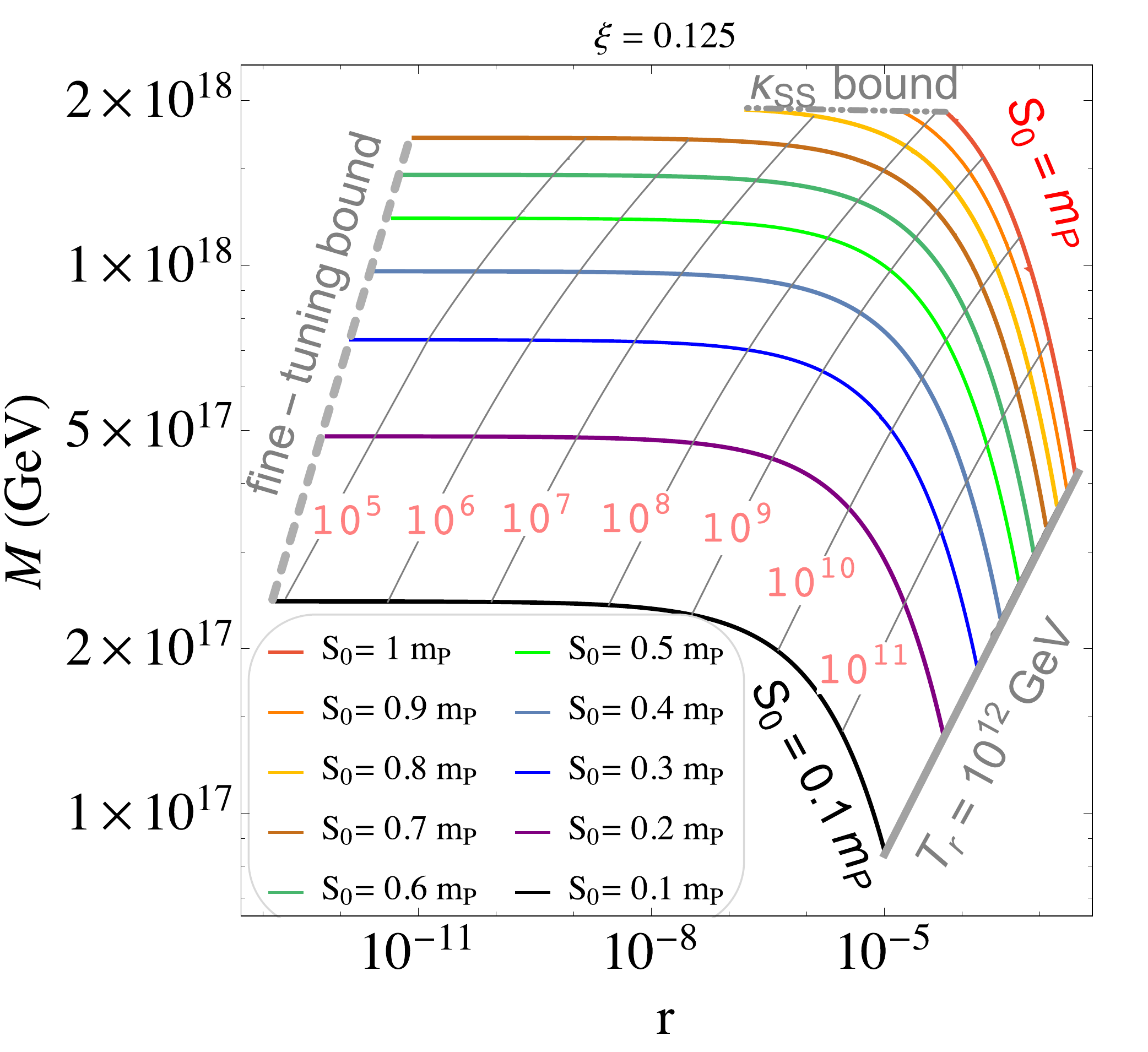}
\par
 \vspace{0.5cm}
\includegraphics[scale=0.425]{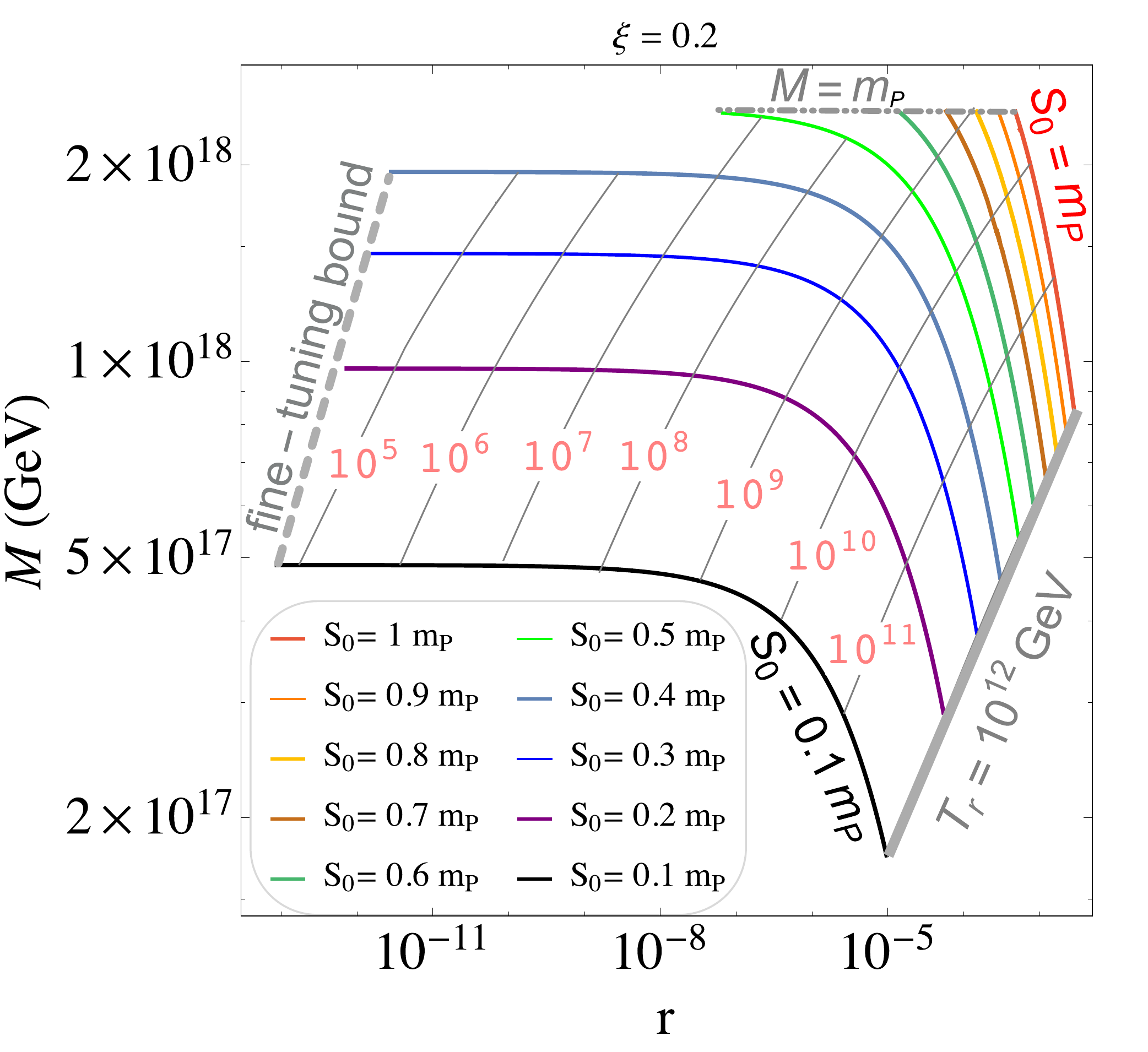}
\caption{\label{fig:larger1} The mass scale $M$ versus the tensor-to-scalar ratio $r$ for 
$\xi=0.125$ and $\xi=0.2$ in the upper and lower plot respectively. The gravitino mass $m_{3/2}\sim1-100$ TeV, $n_s=0.9655$, $\gamma=2$, and $S_0=(0.1-1)\ m_P$. The solid-gray lines are the constant reheat temperature curves
ranging from $10^5-10^{12}~{\rm GeV}$. The dashed-gray line represents the fine-tuning bound, and the double-dot-dashed line represents either the upper bound on $\kappa_{SS}$ or the points where $M=m_P$.} 
\end{figure} 
 \begin{figure}[ht!]
\includegraphics[scale=0.425]{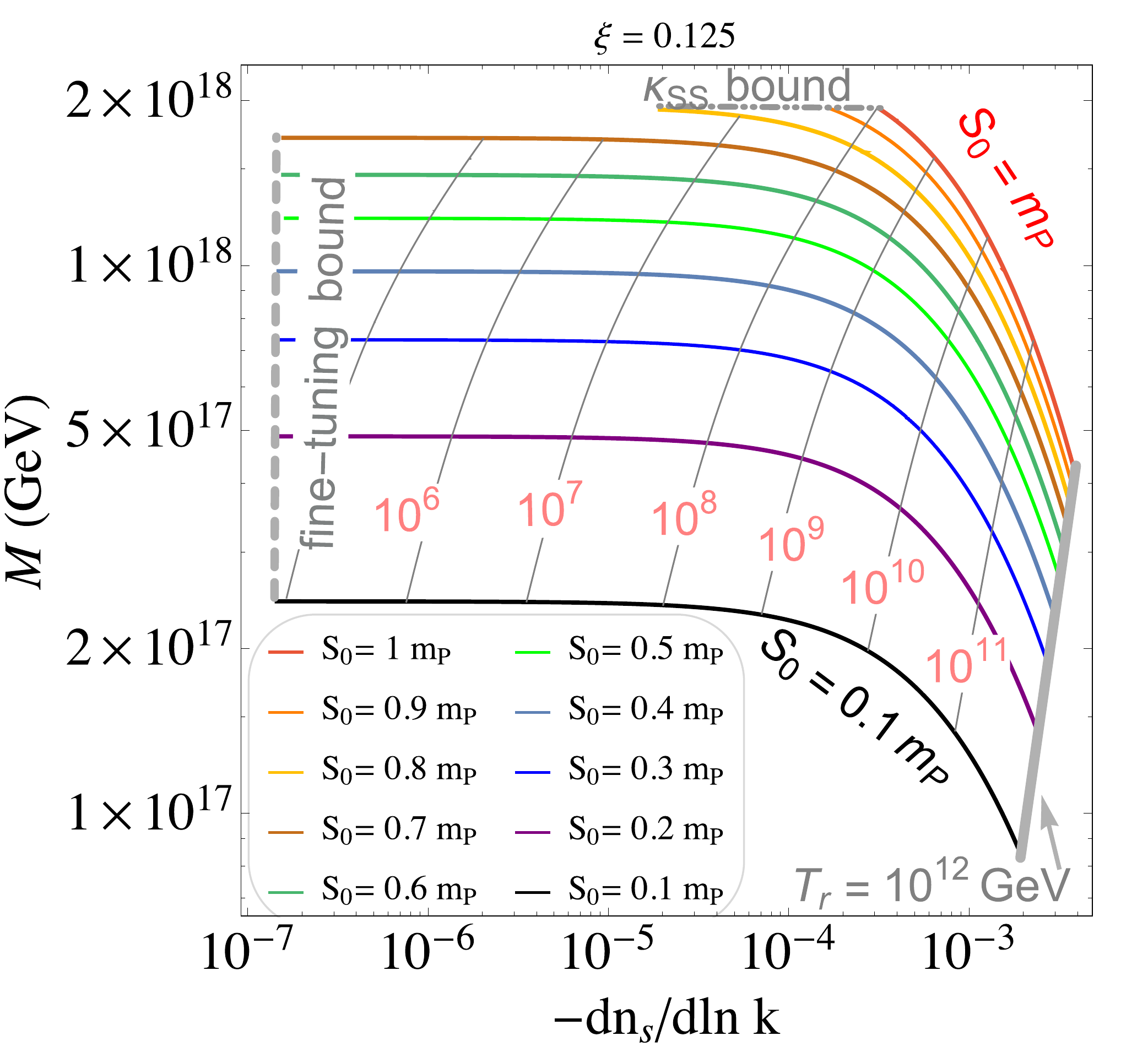}
\par
 \vspace{0.5cm}
\includegraphics[scale=0.425]{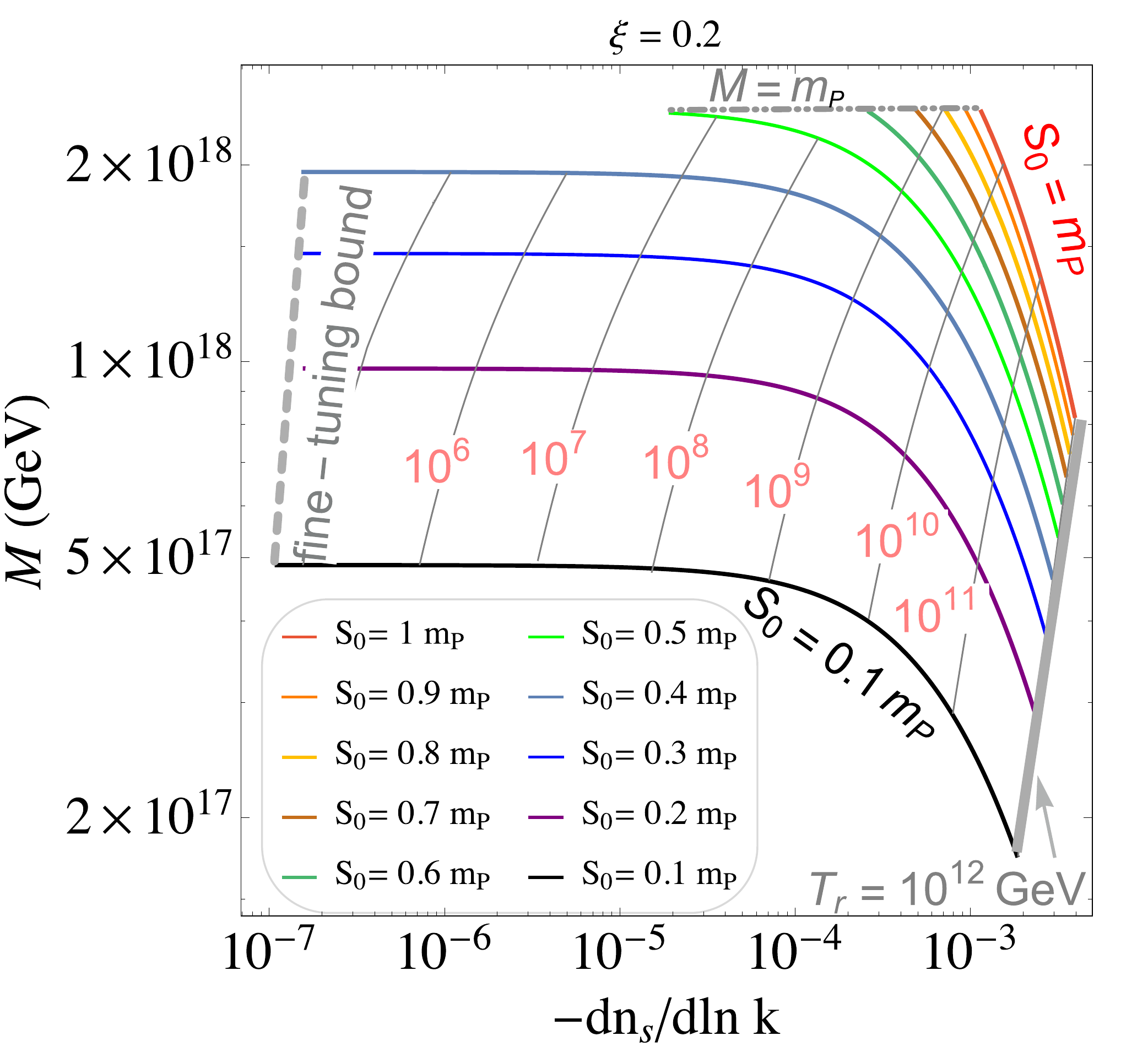}
\caption{\label{fig:larger2} The mass scale $M$ versus the running of the scalar spectral index $dn_s/d\ln k$ for $\xi\!=\!0.125$ and $\xi\!=\!0.2$ in the upper and lower plot respectively. The gravitino mass $m_{3/2}\!\sim\!1\!-\!100$ TeV, $n_s\!=\!0.9655$, $\gamma\!=\!2$, and $S_0=(0.1\!-\!1)\ m_P$. The solid-gray lines are the constant reheat temperature curves ranging from $10^5\!-\!10^{12}$ GeV. The dashed-gray line
shows the fine-tuning bound, and the double-dot-dashed line shows either the upper bound on $\kappa_{SS}$ or the points where $M\!=\!m_P$.} 
\end{figure}
    The canonical measure of primordial gravity waves is the tensor-to-scalar ratio $r$ and the next-generation experiments are gearing up to measure it. One of the highlights of PRISM \cite{Andre:2013afa} is to detect inflationary gravity waves with $r$ as low as $5\times10^{- 4}$ and a major goal of LiteBIRD \cite{Matsumura:2013aja} is to attain a measurement of $r$  within an uncertainty of $\delta r = 0.001$. Future missions include PIXIE \cite{Kogut:2011}, which aims to measure $r < 10^{-3}$ at $5$ standard deviations, and CORE \cite{Finelli:2016cyd}, which forecasts to lower the detection limit for the tensor-to-scalar ratio down to the $10^{-3}$ level. 

\begin{figure}[ht!]
\includegraphics[scale=0.425]{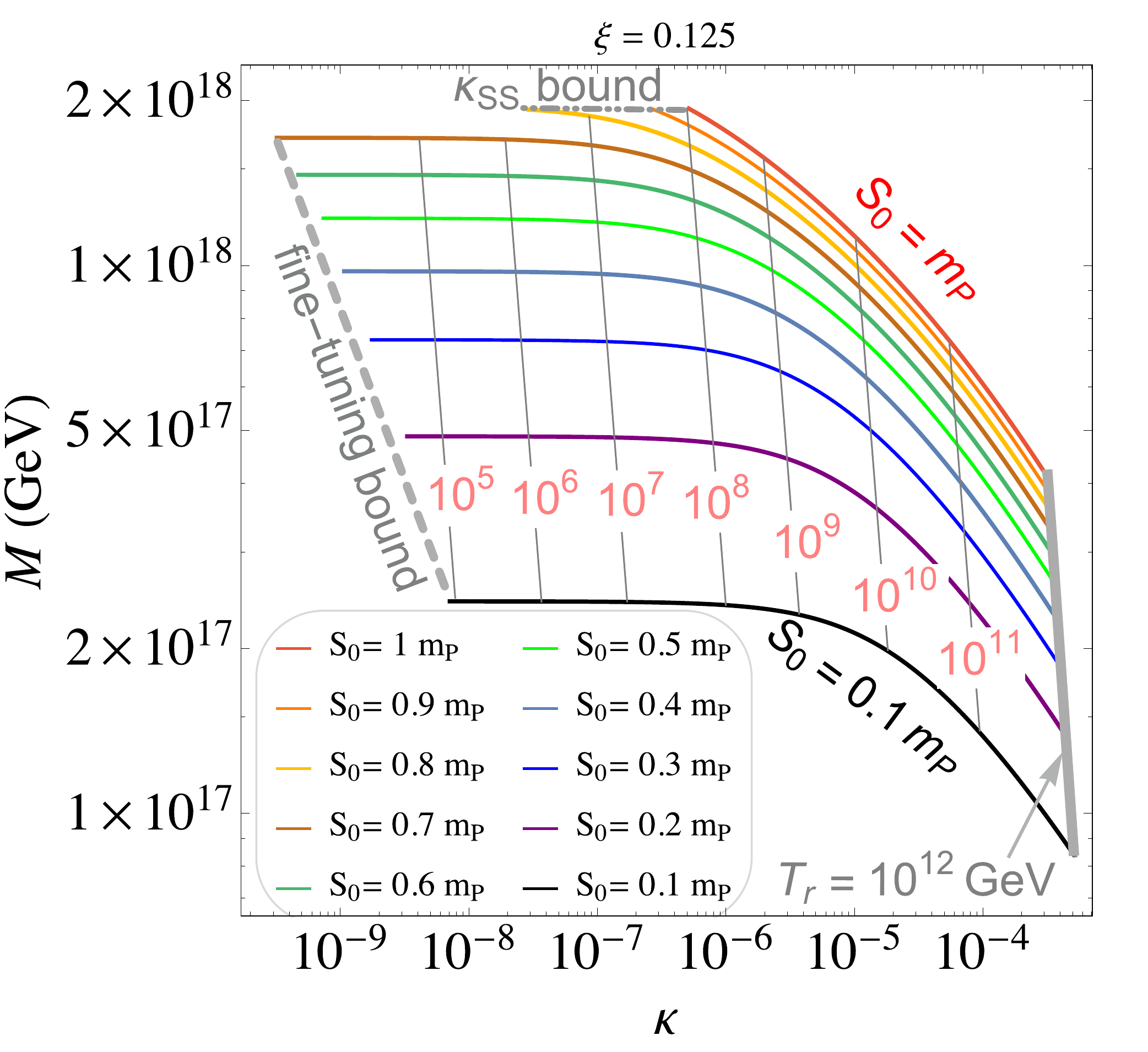}
\par
 \vspace{0.5cm}
\includegraphics[scale=0.425]{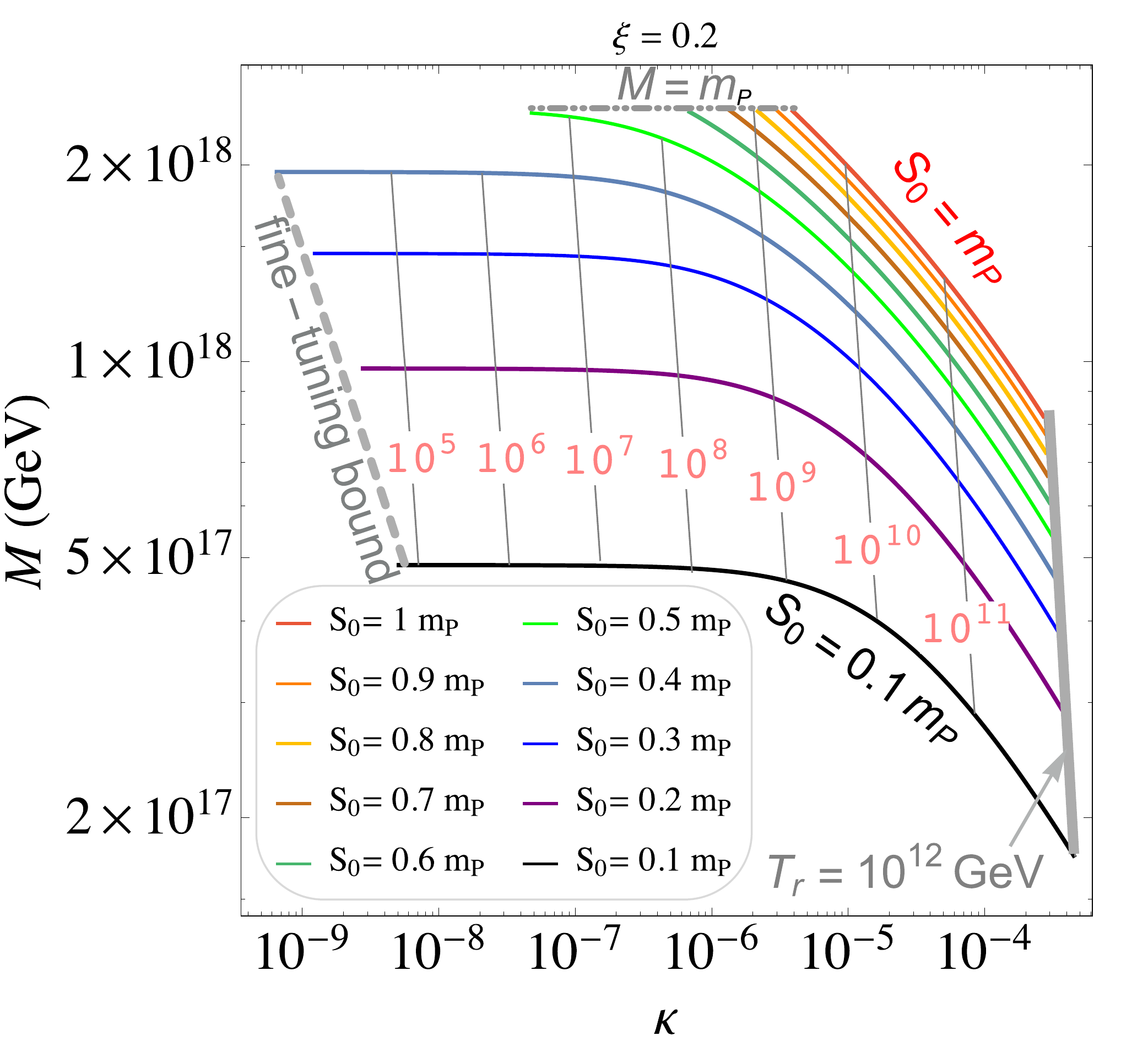}
\caption{\label{fig:larger3} The mass scale $M$ versus $\kappa$ for $\xi=0.125$ and $\xi=0.2$ in the upper and lower plot respectively. The gravitino mass  $m_{3/2}\sim1-100$ TeV, 
$n_s=0.9655$, $\gamma=2$ and $S_0=(0.1-1)\ m_P$. The solid-gray lines are the constant reheat temperature curves
ranging from $10^5-10^{12}~{\rm GeV}$. The dashed-gray line represents the fine-tuning bound, and the double-dot-dashed
line represents either the upper bound on $\kappa_{SS}$ or the points where $M=m_P$.} 
\end{figure}

\begin{figure}
\includegraphics[scale=0.6]{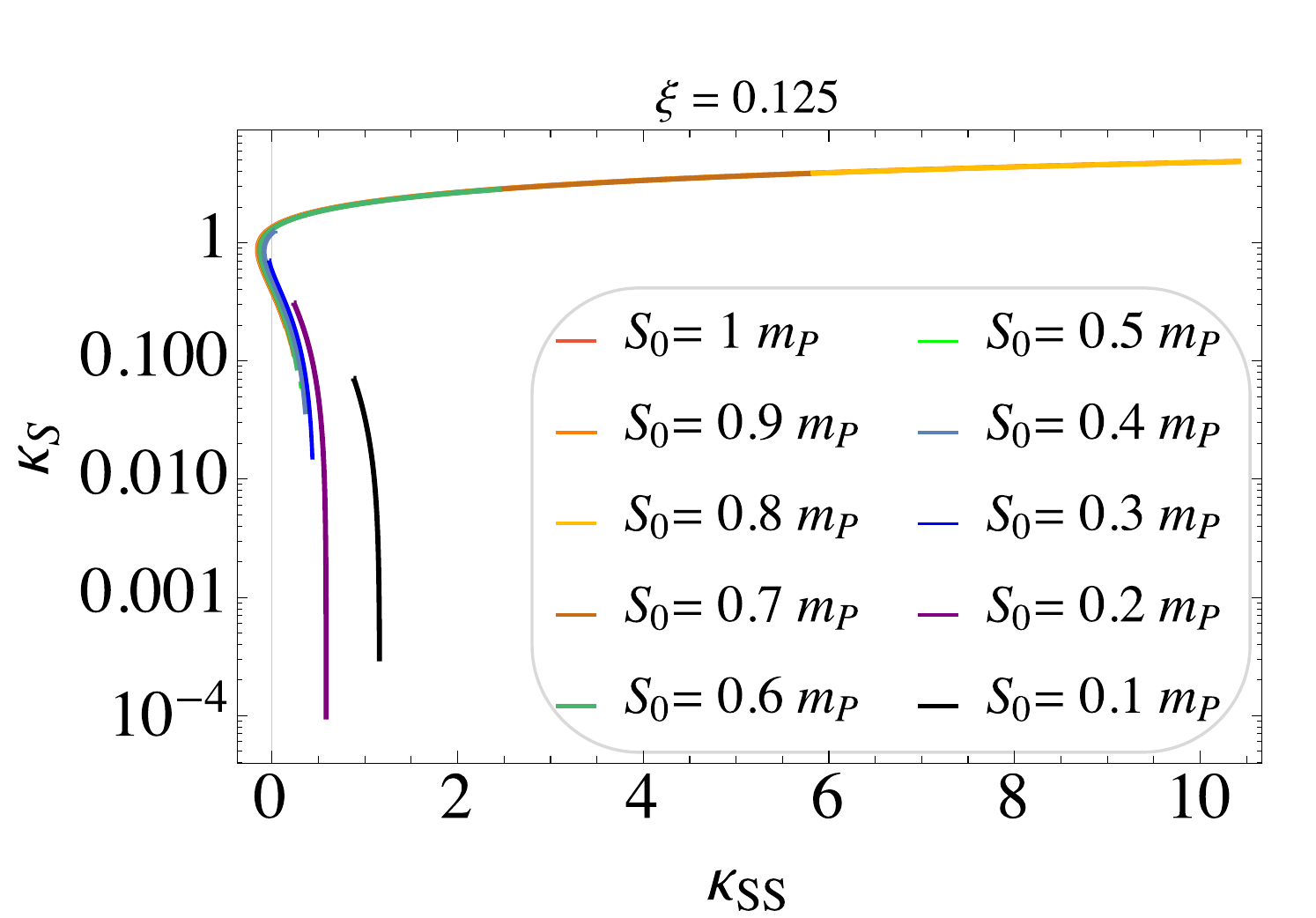}
\par
 \vspace{0.525cm}
\includegraphics[scale=0.6]{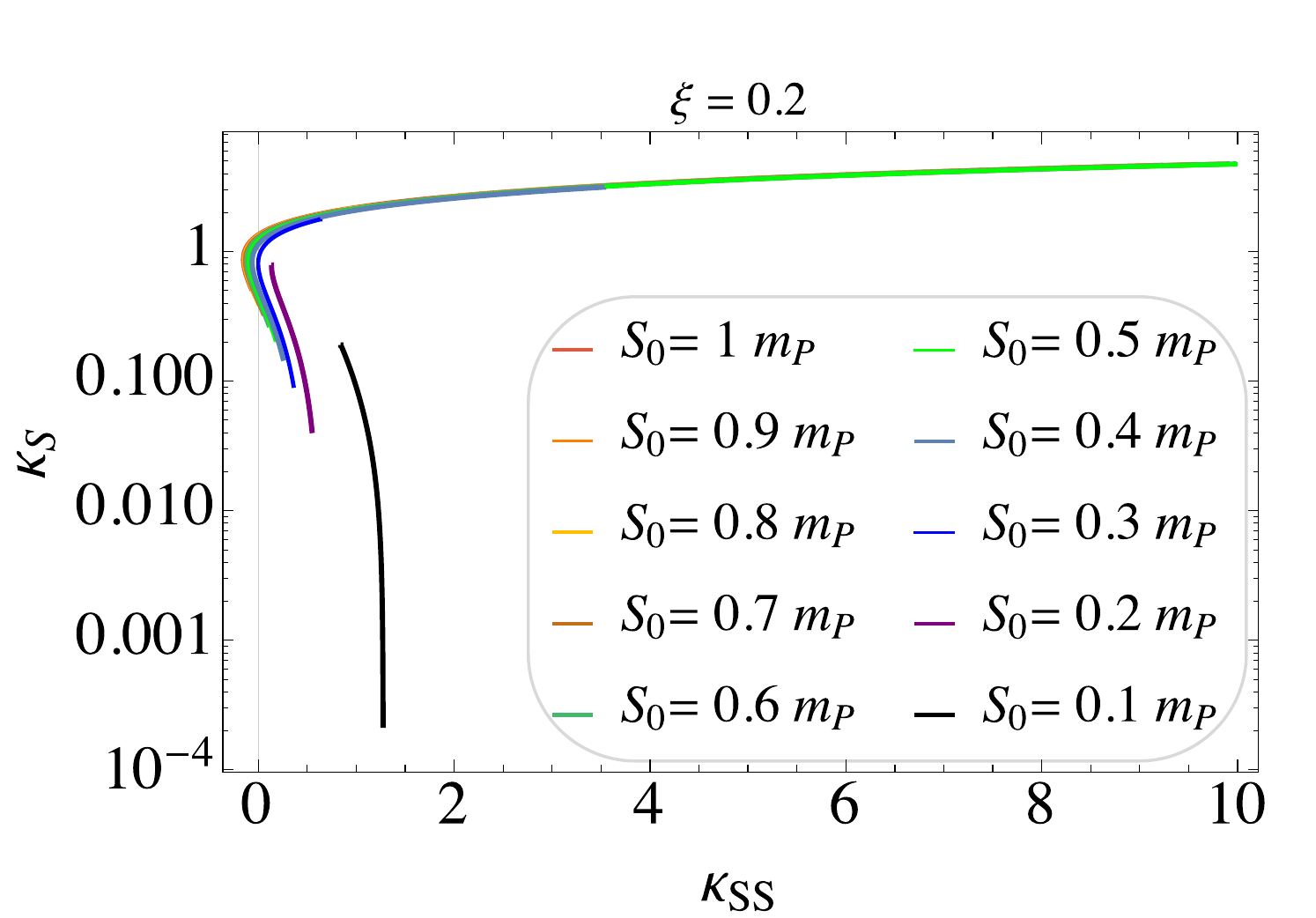}
\caption{\label{fig:larger4} The variation of the couplings $\kappa_S$ and $\kappa_{SS}$ for $\xi=0.125$ and $\xi=0.2$ in the upper and lower plot respectively. The gravitino mass range $m_{3/2}\sim1-100$ TeV, $n_s=0.9655$, $\gamma=2$, and $S_0=(0.1-1)\ m_P$.} 
\end{figure}

 As seen in previous sections, with $\kappa_{SS}=0$, the tensor-to-scalar ratio remains in the undetectable range $r\lesssim 10^{-6}$. It is therefore instructive to explore our model further to look for large-$r$ solutions, which, as it turns out, yield $r$'s in the $10^{-4}-10^{-3}$ range. To achieve this, we employ nonzero $\kappa_{SS}$ in addition to a nonzero $\kappa_S$, and the results are presented in Figs.~\ref{fig:larger1}--\ref{fig:larger4}, for a range of values of the field $S$ at horizon crossing of the pivot scale    
$S_0=(0.1 -1)\ m_P$. In addition, the variation of the parameter $\xi$ is also depicted in these figures by plotting results with $\xi=0.125$ and $\xi=0.2$.

The curves corresponding to field values $S_0$ close to $m_P$ are terminated since, at some point,  either the nonminimal coupling $\big|\kappa_{SS}\big|$ takes unnatural values $\approx 10$ (see Fig.~\ref{fig:larger4}) or $M$ reaches $m_P$. Indeed, for $\xi=0.125$, the coupling $\big|\kappa_{SS}\big|$ can exceed the bound of 10 on curves with $S_0\geq 0.8\ m_P$ and, for $\xi=0.2$, the mass scale $M$ can exceed $m_P$ on curves with $S_0\geq 0.5\ m_P$. We see that the mass scale $M$ is not independent of $\xi$. In fact, as $\xi$ increases from $\xi=0.125$ to $\xi=0.2$, the mass scale $M$ also increases (this is observed in the $\kappa_{SS}=0$ case as well). 
The curves are terminated at their left end due to the fine-tuning bound that we used in the numerical work. The solid-gray lines in Figs.~\ref{fig:larger1}--\ref{fig:larger3} are the constant reheat temperature lines, starting from the upper cutoff at $T_r=10^{12}$ GeV and going down to values as low as $10^4-10^5{\rm\ GeV}$. 

The upper bound on the tensor-to-scalar ratio $r$, as can be read off from Fig.~\ref{fig:larger1}, is $r\lesssim 0.001$ for the choice of the field $S_0=m_P$ and $r\lesssim10^{-5}$ for $S_0\sim 0.1$~$m_P$. Fig.~\ref{fig:larger1} also shows that $r\lesssim10^{-6}-10^{-3}$ from the requirement that the reheat temperature $T_r\lesssim 10^{11}{\rm\ GeV}$ for circumventing the gravitino problem. The running of the scalar spectral index $dn_s/d\ln k$ remains small namely $10^{-7}\lesssim -dn_s/d\ln k \lesssim 4\times10^{-3}$, as shown in Fig.~\ref{fig:larger2}. 
The variation of the mass scale $M$ with $\kappa$ is shown in Fig.~\ref{fig:larger3}, where we find values of the parameter $\kappa$ up to $5\times 10^{-4}$ for large values of  $M$ ($\sim10^{17}-10^{18}{\rm\ GeV}$). The respective variation in the coupling constants $\kappa_S$ and $\kappa_{SS}$ is shown in Fig.~\ref{fig:larger4}. They remain acceptably small and well within the bound $\big|\kappa_S\big|,\ \big|\kappa_{SS}\big|\lesssim1$, for natural values of $S_0=0.5\ m_P$ or less.
 \begin{table*}[ht!]
\caption{\label{tab:bm} Benchmark points for minimal and nonminimal K\"{a}hler potential, for fixed values of $n_s=0.9655$, $A_s(k_0)=2.196\times10^{-9}$, and $\gamma=2$. Column 1 corresponds to a viable scenario for the minimal case where the NLSP is an unstable gravitino decaying into a neutralino LSP before the neutralino freeze-out. Column 2 corresponds to the maximum value of $r$ ($\sim 10^{-9}$) for $\kappa_{SS}=0$, which turns out to be in the unobservable regime. Column 3 shows that reheat temperatures on the order of $10^9$~GeV are easily obtained for $\kappa_{SS}=0$ and mass scales 
$M\sim 10^{16}$~GeV, close to the GUT scale. Columns 4-6 correspond to non-zero $\kappa_{SS}$'s and large field values at horizon crossing of the pivot scale. In this case, the results become independent of the gravitino mass and can be considered valid for $m_{3/2}\sim(1-100)$~TeV.}
\begin{ruledtabular}
\begin{tabular}{ccccccc}
&1&2&3&4&5&6\\
\hline
$\kappa$&$1.1\times 10^{-2}$&$1.2\times10^{-3}$&$9.6\times10^{-6}$&$5.2\times10^{-4}$&$3.1\times10^{-4}$&$3.4\times10^{-8}$\\
$\kappa_S$&0&0.006&0.017&-0.02&0.19&0.19\\
$\kappa_{SS}$&0&0&0&1.3&0.15&0.84\\
$\xi$ &$0.167$&0.125&0.125&0.125&0.125&0.2\\
$m_{3/2}$\,(GeV)&$ {10^{8}}$&$10^3$&$10^5$&$10^3$&$10^3$&$10^3$\\
$S_0\,(m_P)$&0.04&0.005&0.006&0.1&1&0.1\\
$r$&$1.6\times 10^{-7}$&$2.2\times 10^{-9}$&$2.6\times10^{-12}$&$1.1\times10^{-5}$&$2.9\times10^{-3}$&$4.0\times10^{-12}$\\
${|dn_s/d\ln k|}$&$4.3\times 10^{-4}$&$2.0\times 10^{-4}$&$3.5\times10^{-8}$&$2.0\times10^{-3}$&$3.9\times10^{-3}$&$7.5\times10^{-7}$\\
$M$\,(GeV)&$9.2\times10^{15}$&$6.5\times10^{15}$&$1.4\times10^{16}$&$8.3\times10^{16}$&$4.7\times10^{17}$&$4.8\times10^{17}$\\
$T_r$\,(GeV)&$2.8\times10^{13}$&$10^{12}$ &$10^{9}$&$10^{12}$&$10^{12}$&$10^{6}$\\
\end{tabular}
\end{ruledtabular}
\end{table*}
Although the plots presented in Figs.~\ref{fig:larger1}--\ref{fig:larger4} are for gravitino mass $m_{3/2}=1{\rm\ TeV}$, the curves, for these larger $r$ solutions, are independent of the gravitino mass and are valid for a gravitino mass range $m_{3/2}=1-100{\rm\ TeV}$.

Benchmark points for minimal and nonminimal K\"{a}hler potential, for fixed values of $n_s=0.9655$, 
$A_s(k_0)=2.196\times10^{-9}$, and $\gamma=2$, are given in Table~\ref{tab:bm} along with the   corresponding values of the couplings $\kappa$, $\kappa_S$, $\kappa_{SS}$, $\xi$ and the 
tensor-to-scalar ratio $r$, the running of the spectral index ${|dn_s/d\ln k|}$, the mass scale 
$M$, and the reheat temperature $T_r$. A viable scenario for the minimal case is shown in column 1 with an unstable gravitino being the next-to-LSP (NLSP) and decaying into the neutralino LSP before its freeze-out. Column 2 shows that the maximum value of $r$ for $\kappa_{SS}=0$ is $\sim 10^{-9}$, which is too small to be observable. Column 3 shows that reheat temperatures $\sim 10^9$~GeV can be easily obtained for mass scales $M$ around the GUT scale. At large field values $S_0$, the results are shown in columns 4-6 and are more or less independent of the gravitino mass.
\section{\label{conclusion}Conclusion}
We have implemented a version of SUSY hybrid inflation in $SU(4)_c \times SU(2)_L \times SU (2)_R$, a well motivated extension of the SM.  This maximal subgroup of Spin$(10)$ contains electric charge quantization and arises in a variety of string theory constructions. The MSSM $\mu$ term arises, following Dvali, Lazarides, and Shafi, from the coupling of the electroweak doublets to a gauge singlet superfield playing an essential role in inflation, which takes place along a shifted flat direction. The scheme with minimal K\"ahler potential leads to an intermediate scale gravitino mass 
$m_{3/2}\gtrsim 10^8{\rm\ GeV}$ with the gravitino decaying before the freeze out of the LSP neutralinos and with reheat temperature $T_r\gtrsim 10^{13}{\rm\ GeV}$ \cite{Okada:2015vka}. This points towards split SUSY. In the nonminimal K\"ahler case, we have realized successful inflation with reheat temperatures as low as $10^{5}{\rm\ GeV}$. This is favorable for the resolution of the gravitino problem and compatible with a stable LSP and low-scale ($\sim$TeV) SUSY. Compared with standard $\mu$ hybrid inflation \cite{Rehman:2017gkm}, the reheat temperature is lowered by half an order of magnitude and, due to the additional parameter $\xi$, an order of magnitude increase in the spread of $M$ is seen. We have discussed how primordial monopoles are inflated away and provided a framework that predicts the presence of primordial gravity waves with the tensor-to-scalar ratio $r$ in the observable range ($\sim 10^{-4}-10^{-3}$). This is realized with the mass scale M scale approaching values that are comparable to the string scale ($\sim 5\times10^{17}{\rm\ GeV}$) and a gravitino mass lying in the $1-100{\rm\ TeV}$ range. It is worth noting that the inflaton field values do not exceed the Planck scale, which may be an additional desirable feature in view of the swampland conjectures \cite{Vafa:2005ui,Ooguri:2006in}. For a recent discussion and additional references see Ref.~\cite{Vafa:2019evj}. 
\section*{Acknowledgments}
The work of G.L. and Q.S. was supported by the Hellenic Foundation for Research and Innovation (H.F.R.I.) under the ``First call for H.F.R.I. research projects to support faculty members and researchers and the procurement of high-cost research equipment grant'' (Project Number:2251). 


\begin{thebibliography}{99}
\bibitem{Dvali:1994ms} 
  G.R.~Dvali, Q.~Shafi, and R.K.~Schaefer,
  ``Large scale structure and supersymmetric Inflation without fine-tuning,''
  Phys.\ Rev.\ Lett.\  {\bf 73}, 1886 (1994)
  [hep-ph/9406319].

\bibitem{Copeland:1994vg}
  E.J.~Copeland, A.R.~Liddle, D.H.~Lyth, E.D.~Stewart, and D.~Wands,
  ``False vacuum inflation with Einstein gravity,''
  Phys.\ Rev.\ D {\bf 49}, 6410 (1994)
 [astro-ph/9401011].

\bibitem{Senoguz:2004vu}
V.N.~Senoguz and Q.~Shafi,
``Reheat temperature in supersymmetric hybrid inflation models,''
Phys.\ Rev.\ D {\bf 71}, 043514 (2005)
[hep-ph/0412102].

\bibitem{Rehman:2009nq} 
  M.U.~Rehman, Q.~Shafi, and J.R.~Wickman,
  ``Supersymmetric hybrid inflation redux,''
  Phys.\ Lett.\ B {\bf 683}, 191 (2010)
  [hep-ph/0908.3896].

\bibitem{Pallis:2013dxa}
C.~Pallis and Q.~Shafi,
``Update on minimal supersymmetric hybrid inflation in light of PLANCK,''
Phys. Lett. B \textbf{725}, 327 (2013)
[hep-ph/1304.5202].

\bibitem{Buchmuller:2014epa}
  W.~Buchm\"uller, V.~Domcke, K.~Kamada, and K.~Schmitz,
 ``Hybrid inflation in the complex plane,''
  J.\ Cosmol.\ Astropart.\ Phys.\ {\bf 07}, 054 (2014) 
  [astro-ph.CO/1404.1832].

\bibitem{Akrami:2018odb}
  Y.~Akrami {\it et al.} [Planck Collaboration],
  ``Planck 2018 results. X. Constraints on inflation,''
  astro-ph.CO/1807.06211.

\bibitem{BasteroGil:2006cm} 
  M.~Bastero-Gil, S.F.~King, and Q.~Shafi,
  ``Supersymmetric hybrid inflation with nonminimal K\"ahler potential,''
  Phys.\ Lett.\ B {\bf 651}, 345 (2007)
  [hep-ph/0604198].
  
\bibitem{urRehman:2006hu} 
  M.U.~Rehman, V.N.~Senoguz, and Q.~Shafi,
  ``Supersymmetric and smooth hybrid inflation in the light of WMAP3,''
  Phys.\ Rev.\ D {\bf 75}, 043522 (2007)
  [hep-ph/0612023].

\bibitem{Rehman:2010wm}
M.U.~Rehman, Q.~Shafi, and J.R.~Wickman,
``Observable gravity waves from supersymmetric hybrid inflation II,''
Phys.\ Rev.\ D {\bf 83}, 067304 (2011)
[astro-ph.CO/1012.0309].

\bibitem{Civiletti:2014bca}
M.~Civiletti, C.~Pallis, and Q.~Shafi,
``Upper bound on the tensor-to-scalar ratio in GUT-scale supersymmetric hybrid inflation,''
Phys.\ Lett.\ B {\bf 733}, 276 (2014)
[hep-ph/1402.6254].

\bibitem{Jeannerot:2000sv}
  R.~Jeannerot, S.~Khalil, G.~Lazarides, and Q.~Shafi,
  ``Inflation and monopoles in supersymmetric  ${SU(4)_c\times SU(2)_L \times SU(2)_R}$,''
  J.\ High\ Energy\ Phys.\ {\bf 10}, 012 (2000)
  [hep-ph/0002151].
  
\bibitem{Jeannerot:2001xd}
  G.~Lazarides, ``Supersymmetric hybrid inflation,'' 
  NATO\ Sci.\ Ser.\ II {\bf 34}, 399 (2001) [hep-ph/0011130];
  R.~Jeannerot, S.~Khalil, and G.~Lazarides,
  ``Monopole problem and extensions of supersymmetric hybrid inflation,''  
	Proc.\ of\ CICHEP\ 2001, 254 (2001) [hep-ph/0106035].
  

\bibitem{Pati:1974yy}
J.C.~Pati and A.~Salam,
``Lepton number as the fourth color,''
Phys.\ Rev.\ D {\bf 10}, 275 (1974),
Erratum: Phys.\ Rev.\ D {\bf 11}, 703 (1975).  

\bibitem{King:1997ia}
  S.F.~King and Q.~Shafi,
  ``Minimal supersymmetric $SU(4)_c\times SU(2)_L \times SU(2)_R$,''
  Phys.\ Lett.\ B {\bf 422}, 135 (1998)
  [hep-ph/9711288].

\bibitem{Dvali:1997uq}
  G.R.~Dvali, G.~Lazarides, and Q.~Shafi,
  ``$\mu$ problem and hybrid inflation in supersymmetric $SU(2)_L \times SU(2)_R \times U(1)_{B-L}$,''
  Phys.\ Lett.\ B {\bf 424}, 259 (1998)
  [hep-ph/9710314].

 
\bibitem{Ellis:1984eq} 
  M.Y.~Khlopov and A.D.~Linde,
  ``Is it easy to save the gravitino?,''
  Phys.\ Lett.\ B {\bf 138}, 265 (1984);  J.R.~Ellis, J.E.~Kim, and D.V.~Nanopoulos,
  ``Cosmological gravitino regeneration and decay,''
  Phys.\ Lett.\ B {\bf 145}, 181 (1984).
  
\bibitem{Lazarides:1998qx}
G.~Lazarides and N.~D.~Vlachos,
``Atmospheric neutrino anomaly and supersymmetric inflation,''
Phys. Lett. B \textbf{441}, 46 (1998)
[hep-ph/9807253].


\bibitem{Okada:2015vka}
N.~Okada and Q.~Shafi,
``$\mu$-term hybrid inflation and split supersymmetry,''
Phys.\ Lett.\ B {\bf 775}, 348 (2017)
[hep-ph/1506.01410].

\bibitem{Shafi:1998yy}
  Q.~Shafi and Z.~Tavartkiladze,
  ``Neutrino oscillations and other key issues in supersymmetric $SU(4)_c\times SU(2)_L\times SU(2)_R$,''
  Nucl.\ Phys.\ B {\bf 549}, 3 (1999)
  [hep-ph/9811282].
	
\bibitem{lightdoublets} 
G.~Lazarides and C.~Panagiotakopoulos, ``MSSM from SUSY trinification,''
Phys.\ Lett.\ B {\bf 336}, 190 (1994) [hep-ph/9403317]; 
G.~Dvali and Q.~Shafi,
``Gauge hierarchy, Planck scale corrections and the origin of GUT scale in supersymmetric $(SU(3))^3$,''
Phys. Lett. B \textbf{339}, 241 (1994)
[hep-ph/9404334].

\bibitem{Rehman:2017gkm}
  M.U.~Rehman, Q.~Shafi, and F.K.~Vardag,
  ``$\mu$-hybrid inflation with low reheat temperature and observable gravity waves,''
  Phys.\ Rev.\ D {\bf 96}, 063527 (2017) 
  [hep-ph/1705.03693].

\bibitem{Chamseddine:1982jx}
  A.H.~Chamseddine, R.L.~Arnowitt, and P.~Nath,
  ``Locally supersymmetric grand unification,''
  Phys.\ Rev.\ Lett.\  {\bf 49}, 970 (1982).

\bibitem{Linde:1997sj}
  A.D.~Linde and A.~Riotto,
  ``Hybrid inflation in supergravity,''
  Phys.\ Rev.\ D {\bf 56}, R1841 (1997)
  [hep-ph/9703209].
  
\bibitem{Kyae:2005vg}
B.~Kyae and Q.~Shafi,
``Inflation with realistic supersymmetric SO(10),''
Phys. Rev. D \textbf{72}, 063515 (2005)
[hep-ph/0504044].
  
\bibitem{Coleman;1973}
  S.~Coleman and E.~Weinberg,
  ``Radiative corrections as the origin of spontaneous symmetry breaking,''
  Phys.\ Rev.\ D {\bf 7}, 1888 (1973).
  
\bibitem{Lazarides:1996dv}
G.~Lazarides, R.~K.~Schaefer, and Q.~Shafi,
``Supersymmetric inflation with constraints on superheavy neutrino masses,''
Phys. Rev. D \textbf{56}, 1324 (1997)
[hep-ph/9608256].
  
    
\bibitem{Panagiotakopoulos:1997ej}
C.~Panagiotakopoulos,
``Hybrid inflation with quasicanonical supergravity,''
Phys. Lett. B \textbf{402}, 257 (1997)
[hep-ph/9703443].
  

\bibitem{Lazarides:1998zf}
G.~Lazarides and N.~Tetradis,
``Two stage inflation in supergravity,''
Phys. Rev. D \textbf{58}, 123502 (1998)
[hep-ph/9802242].

\bibitem{Dimopoulos:2011ym}
K.~Dimopoulos, G.~Lazarides, and J.~M.~Wagstaff,
``Eliminating the $\eta$-problem in SUGRA Hybrid Inflation with Vector Backreaction,''
JCAP \textbf{02}, 018  (2012)
[astro-ph.CO/1111.1929].

\bibitem{Haba:2005ux}
N.~Haba and N.~Okada,
``Structure of split supersymmetry and simple models,''
Prog. Theor. Phys. \textbf{114}, 1057 (2006)
[hep-ph/0502213].

\bibitem{Wu:2016fzp}
L.~Wu, S.~Hu, and T.~Li,
``No-Scale $\mu$-Term Hybrid Inflation,''
Eur. Phys. J. C \textbf{77} no.3, 168 (2017) 
[hep-ph/1605.00735].

\bibitem{Civiletti:2011qg}
  M.~Civiletti, M.U.~Rehman, Q.~Shafi, and J.R.~Wickman,
  ``Red spectral tilt and observable gravity waves in shifted hybrid inflation,''
  Phys.\ Rev.\ D {\bf 84}, 103505 (2011) 
  [astro-ph.CO/1104.4143].


\bibitem{Rehman:2009yj}
  M.U.~Rehman, Q.~Shafi, and J.R.~Wickman,
  ``Minimal supersymmetric hybrid inflation, flipped SU(5) and proton decay,''
  Phys.\ Lett.\ B {\bf 688}, 75 (2010)
  [hep-ph/0912.4737].
 
\bibitem{Bolz:2000fu} 
  M.~Bolz, A.~Brandenburg, and W.~Buchmuller,
  Thermal production of gravitinos,
  Nucl.\ Phys. {\bf  B606}, 518 (2001); Nucl.\ Phys. {\bf B790}, 336(E) (2008)
 [hep-ph/0012052].
  
\bibitem{Tanabashi:2018oca}
M.~Tanabashi \textit{et al.} [Particle Data Group],
``Review of Particle Physics,''
Phys. Rev. D \textbf{98} no.3, 030001(2018). 
  

\bibitem{Khlopov:1993ye}
M.~Khlopov, Y.~Levitan, E.~Sedelnikov, and I.~Sobol,
``Nonequilibrium cosmological nucleosynthesis of light elements: Calculations by the Monte Carlo method,''
Phys.\ Atom.\ Nucl.\ {\bf 57}, 1393 (1994).

\bibitem{Kawasaki:2004qu}
  M.~Kawasaki, K.~Kohri, and T.~Moroi,
  ``Big-Bang nucleosynthesis and hadronic decay of long-lived massive particles,''
  Phys.\ Rev.\ D {\bf 71}, 083502 (2005) 
  [astro-ph/0408426].

\bibitem{Kawasaki:2017bqm}
  M.~Kawasaki, K.~Kohri, T.~Moroi, and Y.~Takaesu,
  ``Revisiting big-bang nucleosynthesis constraints on long-lived decaying particles,''
  Phys.\ Rev.\ D {\bf 97}, 023502 (2018)  
  [hep-ph/1709.01211].
  
\bibitem{Hooper:2002nq} 
  D.~Hooper and T.~Plehn,
 ``Supersymmetric dark matter: How light can the LSP be?,''
  Phys.\ Lett.\ B {\bf 562}, 18 (2003)
  [hep-ph/0212226].

\bibitem{Kawasaki:2008qe} 
  M.~Kawasaki, K.~Kohri, T.~Moroi, and A.~Yotsuyanagi,
 ``Big-bang nucleosynthesis and gravitino,''
  Phys.\ Rev.\ D {\bf 78}, 065011 (2008)
  [hep-ph/0804.3745].
  
\bibitem{Khlopov:1984pf}
M.~Khlopov and A.D.~Linde,
``Is it easy to save the gravitino?,''
Phys.\ Lett.\ B {\bf 138}, 265 (1984).
 
\bibitem{Okada:2017rbf}
N.~Okada and Q.~Shafi;
``Gravity waves and gravitino dark matter in $\mu$-hybrid inflation,''
Phys. Lett. B \textbf{787}, 141 (2018)
[hep-ph/1709.04610].
 
\bibitem{Andre:2013afa}
  P.~Andre {\it et al.} (PRISM Collaboration),
  ``PRISM (Polarized Radiation Imaging and Spectroscopy Mission): A white paper on the ultimate polarimetric spectro-imaging of the microwave and far-infrared sky,''
 astro-ph.CO/1306.2259.
  
\bibitem{Matsumura:2013aja}
  T.~Matsumura {\it et al.},
  ``Mission design of LiteBIRD,''
  J.\ Low.\ Temp.\ Phys.\  {\bf 176}, 733 (2014) 
  [astro-ph.CO/1311.2847].

\bibitem{Kogut:2011}
A.~Kogut {\it et al.},
 ``The Primordial Inflation Explorer (PIXIE): a nulling polarimeter for cosmic microwave background observations,''
  J.\ Cosmol.\ Astropart.\ Phys. {\bf 07}, 025 (2011)
 [astro-ph.CO/1105.2044].
 
\bibitem{Finelli:2016cyd}
  F.~Finelli {\it et al.} [CORE Collaboration],
  ``Exploring cosmic origins with CORE: Inflation,''
  JCAP {\bf 04}, 016 (2018)
  [astro-ph.CO/1612.08270].

\bibitem{Vafa:2005ui}
C.~Vafa,
``The string landscape and the swampland,''
hep-th/0509212.

\bibitem{Ooguri:2006in}
H.~Ooguri and C.~Vafa,
``On the geometry of the string landscape and the swampland,''
Nucl.\ Phys.\ B {\bf 766}, 21 (2007)
[hep-th/0605264].
\bibitem{Vafa:2019evj}
C.~Vafa,
``Cosmic Predictions from the String Swampland,''
APS Physics \textbf{12}, 115 (2019).

\end{thebibliography}
\end{document}